\documentclass[journal]{IEEEtran}
\usepackage{amsfonts}
\usepackage{amssymb}
\usepackage{amsmath}
\usepackage{amsthm}
\usepackage{algorithm}
\usepackage{array}
\usepackage{algpseudocode}
\usepackage{changepage}
\usepackage{color}
\usepackage[compress]{cite}
\usepackage{epstopdf}
\usepackage{graphicx}
\usepackage{multirow}
\usepackage{subfigure}
\usepackage{setspace}
\usepackage{xfrac}	
\usepackage[table,xcdraw]{xcolor}	

\def\ps@headings{%
\def\@oddhead{\mbox{}\scriptsize\rightmark \hfil \thepage}%
\def\@evenhead{\scriptsize\thepage \hfil \leftmark\mbox{}}%
\def\@oddfoot{}%
\def\@evenfoot{}}

\newtheorem{thm}{\textbf{Theorem}\bfseries}			
\newtheorem{lem}[thm]{\textbf{Lemma}\bfseries}		
\newtheorem{corl}[thm]{\textbf{Corollary}\bfseries}		
 \newtheorem{pro}[thm]{\textbf{Property}\bfseries}		

\renewcommand{\baselinestretch}{0.915}

\begin{document}
\title{\LARGE DTER: Schedule Optimal RF Energy Request and Harvest for \\Internet of Things}

\author{Yu Luo, ~\IEEEmembership{Member, IEEE}, Lina Pu, ~\IEEEmembership{Member, IEEE}, Yanxiao Zhao,~\IEEEmembership{Member, IEEE},  Guodong Wang\\ and Min Song,~\IEEEmembership{Senior Member, IEEE}
\thanks{Y. Luo, L. Pu, Y. Zhao, and G. Wang are with South Dakota School of Mines and Technology. Email: \{yu.luo, lina.pu, yanxiao.zhao, guodong.wang\}@sdsmt.edu. M. Song is with Michigan Technological University. Email: mins@mtu.edu.}}

\maketitle

\begin{abstract}
\label{sec:abstract}
We propose a new energy harvesting strategy that uses a dedicated energy source (ES) to optimally replenish energy for radio frequency (RF) energy harvesting powered Internet of Things. Specifically, we develop a two-step dual tunnel energy requesting (DTER) strategy that minimizes the energy consumption on both the energy harvesting device and the ES. Besides the causality and capacity constraints that are investigated in the existing approaches, DTER also takes into account the overhead issue and the nonlinear charge characteristics of an energy storage component to make the proposed strategy practical. Both offline and online scenarios are considered in the second step of DTER. To solve the nonlinear optimization problem of the offline scenario, we convert the design of offline optimal energy requesting problem into a classic shortest path problem and thus a global optimal solution can be obtained through dynamic programming (DP) algorithms. The online suboptimal transmission strategy is developed as well. Simulation study verifies that the online strategy can achieve almost the same energy efficiency as the global optimal solution in the long term.
\end{abstract}

\begin{IEEEkeywords}
Energy harvesting for Internet of Things, Radio frequency (RF) energy harvest, Schedule optimal energy request.
\end{IEEEkeywords} 
\section{Introduction}
\label{sec:Introduction}

With the worldwide progress toward Internet of Things (IoT), the number of sensors deployed and connected to the Internet is growing at a rapid pace~\cite{gubbi2013internet}. Energy harvesting (EH) has been considered as a favorable supplement to drive the numerous sensors in the emerging IoT. Due to several key advantages like the pollution free, long lifetime, and energy self-sustainability, the EH-IoT systems are competitive in a wide spectrum of applications (e.g., healthcare, surveillance, and emergency response to  natural and man-made disasters)~\cite{kamalinejad2015wireless}. With the EH technique, an energy harvesting device (EHD) is capable of receiving energy from nature or man-made sources, such as solar, wind or radio frequency (RF) signals, for communications~\cite{kazmierski2014energy}. Along with the remarkable progress on ultra-low power semiconductors, RF-based EH technique has drawn considerable attention in recent years~\cite{ulukus2015energy, zhou2015greendelivery}.

The existing RF-based EH systems can be classified into two categories: \emph{without a dedicated energy source} (ES) and \emph{with a dedicated ES}. In the first category, EHDs harvest energy passively from ambient RF signal, which are radiated from a TV tower, an access point (AP), or a base station~\cite{pinuela2013ambient, liu2013ambient}. The performance of a system in this category could be improved by optimizing the data rate and transmission power of EHDs. The activities of ESs, however, are uncontrollable due to a lack of interaction between EHDs and ESs. In the second category, a dedicated ES is installed to power the entire EH network~\cite{luo2017optimal, mishra2015smart, kaushik2013experimental, huang2014enabling}. In this category, EHDs are allowed to request energy from the associated ES on demand. The replenishment of energy storage components, which is also referred to as battery charging, thus can be scheduled by the ES based on the energy consumption of each EHD.

Current research on RF-based EH communications mainly focuses on power management, to maximize the utilization of the harvested energy from RF ambient signal~\cite{yang2012optimal, tutuncuoglu2012optimum, lu2015resource}. The investigation on EH communications with a dedicated ES, by contrast, is still in its initial stage. Actually, the EH systems with a dedicated ES like passive RF identification (RFID) have already been penetrating our daily lives~\cite{das2014rfid, kaur2011rfid}. A number of advanced power supply methods, such as multi-hop energy packet transmission and sharp beamforming energy transfer, are also investigated or tested through theoretical analysis and experiments~\cite{percy2012supplying, huang2015cutting, mishra2015smart, kaushik2013experimental}. How to manage the power at EHDs and schedule the energy transfer at ESs to minimize the overall energy consumption, however, is currently missing and is the focus of this paper.

Intuitively, an RF-based EH system could achieve a satisfactory performance in terms of transmission rate, packet loss and reliability by allowing an EHD to request the energy freely from a dedicated ES. The energy efficiency with such a greedy strategy, however, will be reduced considerably. Specifically, due to the overhead on requesting energy and the nonlinear charging feature of batteries, the energy consumption of an ES system varies dramatically when different energy harvesting strategies are applied. As will be revealed in this paper, if an inappropriate strategy is employed, a large amount of energy is wasted either on sending superfluous request messages or on an inefficient charging process. How to minimize the overall energy consumption while guaranteeing efficient data transmission for wireless communications remains a challenge.

In this paper, we propose a two-step dual tunnel energy requesting (DTER) strategy to minimize the energy consumption at both the EHD and the ES on timely data transmission. 
The proposed strategy is operated in two steps: the first step derives the profile of optimal transmission rate in a feasible \emph{data tunnel} at an EHD under a data storage constraint; the second step designs an offline global optimal strategy for energy harvesting in a feasible \emph{energy tunnel} based on the first step subject to the constraint of limited battery capacity. By using DTER, an EHD is able to harvest energy from a dedicated ES timely and efficiently with the minimum energy consumption on both EHD and ES.

To make the proposed strategy more practical, we also develop an online suboptimal strategy for the second step of DTER. In the online strategy, the EHD just needs to monitor the residual energy in its battery and requests a certain amount of energy once the residual energy drops to a predefined threshold. Theoretical analysis verifies that in the suboptimal strategy, if the charging rate of an energy storage component is much higher than the EHD's transmission power, both the energy requested by an EHD and that left at the energy storage are constants, which makes the online strategy easy to implement in a real system. According to simulation results, the energy efficient of the online transmission strategy is close to the global optimal solution. 


The remainder of the paper is organized as follows. We introduce the related work in Section~\ref{sec:RelatedWork} and the system model in Section~\ref{sec:system}. Section~\ref{sec:optTranRate} discusses the first step of DTER to obtain the optimal transmission rate of an EHD. Two important concepts used in the second step of DTER are described in Section~\ref{sec:PreReq}. We study how to design an offline global optimal energy requesting strategy to minimize the overall energy consumption in Section~\ref{sec:EngReq}. An online suboptimal strategy is presented Section~\ref{sec:OnlineOpt}. Section~\ref{sec:PerEval} presents simulation results. Conclusions are drawn in Section~\ref{sec:conclusion}. For the convenience of reading the manuscript, we list the definitions of all symbols used in Appendix~\ref{app:appB}.

\section{Related Work}
\label{sec:RelatedWork}

As a promising technique to build the emerging green and self-sustainable IoT, the EH technique has been extensively investigated. Many policies are proposed for EH without a dedicated ES to manage the harvested energy efficiently after considering the intermittency and the randomness of energy arrivals~\cite{yang2012optimal, tutuncuoglu2012optimum, lu2015resource}.

Reference \cite{yang2012optimal} introduces an optimal power control policy that minimizes the completion time of transmitting a certain amount of data under an energy storage constraint. In \cite{tutuncuoglu2012optimum}, both the constraints of energy and data storage capacities are taken into account to maximize the short-term throughput of an EHD. The optimal transmission power is shown to be the tightest string that lies in a feasible energy tunnel. The authors in \cite{lu2015resource} propose an optimal operation strategy that provides service differentiation among different traffic patterns, subject to the constraints of data storage capacity and packet loss ratio. The activity of an EHD is modeled as a constrained Markov decision process and the optimal decision on whether to harvest energy or to send data is figured out.

Although harvesting ambient RF energy solely from surrounding environments to power EHDs is attractive, recent research studies have demonstrated that the transmission rate, reliability, transmission range, and deployment of such devices are extremely limited by the thin energy in the atmosphere~\cite{lu2015wireless, liu2013ambient, pinuela2013ambient}. To meet communication requirements (e.g., reliability, delay, and throughput), powering EHDs with a dedicated ES is promoted as an alternative. 

Several experiments are conducted in \cite{mishra2015smart} and \cite{kaushik2013experimental} showing that multi-hop RF energy transfer can save energy on ES and reduce the time consumption on energy harvests as well. Reference \cite{percy2012supplying} proposes a system for power transfer on an autonomous radio frequency. It is capable of rotating the base of ES to track the position of EHDs and transferring power to a particular device. In \cite{huang2015cutting}, the authors advocate the use of massive multiple-input and multiple-output (MIMO) to enable a sharp beamforming for efficient wireless energy transfer. The authors in~\cite{naderi2014rf} propose a CSMA/CA based medium access control protocol to schedule both the energy transmission and data communication in shared medium; the problem of ES deployment is also explored in this work.

From the above introduction we realize that the feasibility, hardware design and applications of an EH system with a dedicated ES have been comprehensively studied in the literature; how to efficiently manage the power at both the ES and the EHD, however, is still an open issue. As will be shown in the paper, inappropriate energy requests at the EHD may significantly increase the energy consumption at the ES.  Therefore, the objective of this paper is to design an optimal energy requesting strategy that the EHD could harvest energy from the ES with the minimum energy consumption.

\section{System Model}
\label{sec:system}

\begin{figure}[htb]
\centerline{\includegraphics[width=6.8cm]{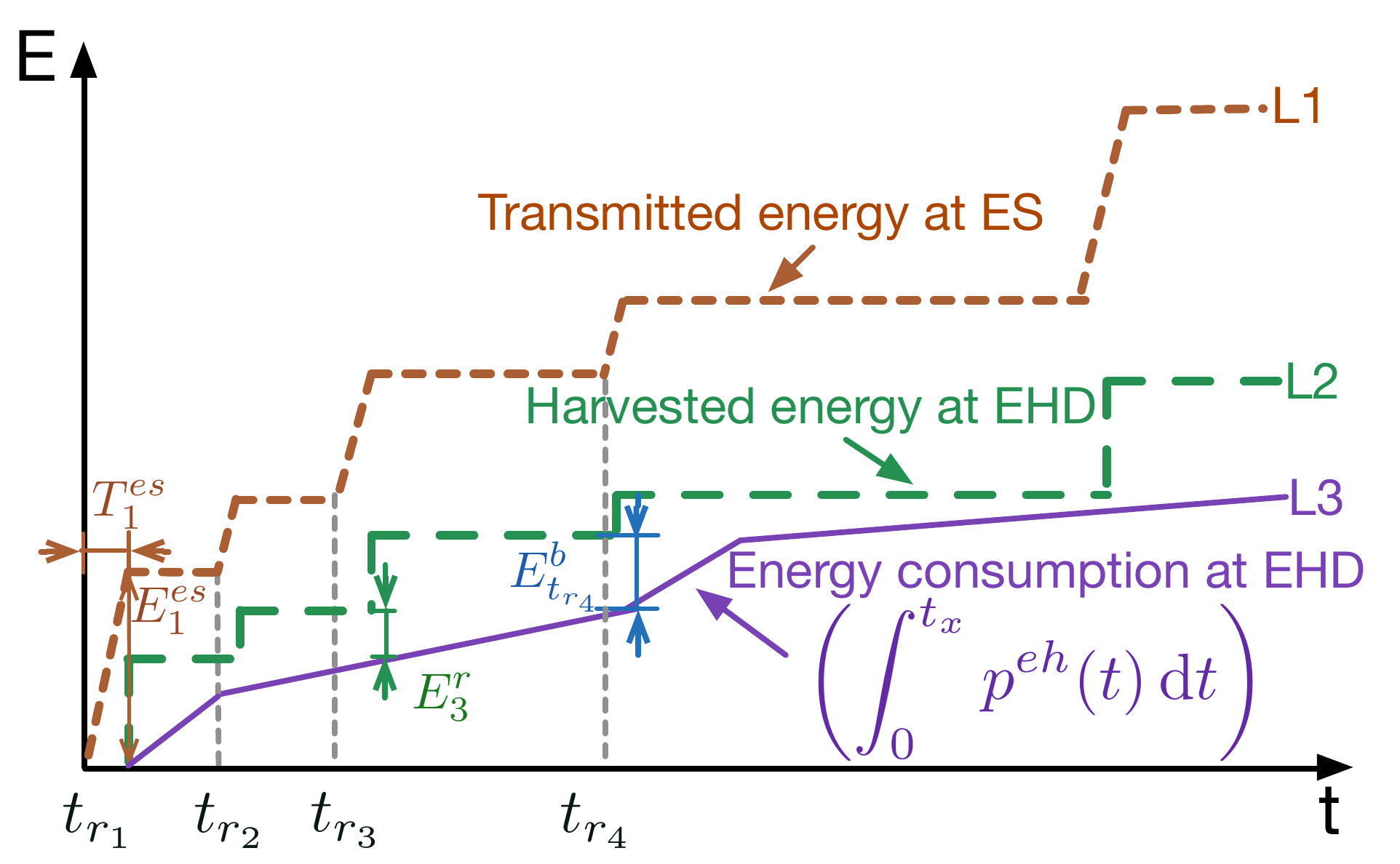}}
\caption{System model of energy request.}\label{fig:EnergyRequest}
\end{figure}

We consider an RF-based EH system with a dedicated ES, from which an EHD requests and harvests energy through RF energy packets. 
Assume the energy harvests and data transmissions use separate antennas and frequency bands, i.e., an EHD can request energy and transmit data simultaneously.
Fig.~\!\ref{fig:EnergyRequest} illustrates the energy request problem. Three curves, labeled by L1, L2, and L3, represent the cumulative transmitted energy by the ES, the accumulation of harvested energy and energy consumption at the EHD, respectively. The harvested energy is considered as discrete energy packets with different sizes~\!\cite{yang2012optimal, tutuncuoglu2012optimum, lu2015resource}. Therefore the profile of harvested energy at the EHD (i.e., L2) is modeled as a staircase curve. As shown in Fig.~\!\ref{fig:EnergyRequest}, the EHD initiates the $i^{th}$ energy request at instance $t_{r_i}$ and harvests $E_i^r$ energy in the next $T_i^{es}$ seconds. Here, $T_i^{es}$ is the duration of the energy packet, which is also the charge time for the $i^{th}$ energy request. Assume the ES consumes $E_i^{es}$ energy on transmitting the $i^{th}$ energy packet. Due to the energy loss on propagation and charging, the harvested energy $E_i^r$ is always smaller than the transmitted energy $E_i^{es}$.

As for data transmissions, we consider both offline and online scenarios, where offline/online assumes that the arriving time ($t_{d_n}$) and the size ($D_n$) of future data are known/unknown exactly by the EHD. The offline scenario is studied first and the online strategy will be discussed in Section~\ref{sec:OnlineOpt}.
 Let $D_m^{eh}$ be the maximum capacity of data storage and $D_t^b$ be the residual data at instance $t$. The transmission rate of the EHD at time $t$ is represented by $r^{eh}(t)$, which is related to the transmission power, $p^{eh}(t)$, through a power-rate function $r^{eh}(t)=g\left(p^{eh}(t)\right)$\footnote{A typical power-rate function in an additive white Gaussian noise (AWGN) channel is $r^{eh}(t)\!=\!\log[1+h(t)p^{eh}(t)]$, where $h(t)$ is the instantaneous channel response for the link between an EHD and its communicating party.}. An illustration of the cumulative energy consumption on data transmission, $\int_0^{t_x}p^{eh}(t)\,\mathrm{d}t$, labeled by L3, is shown in Fig.~\!\ref{fig:EnergyRequest}.

Let $T$ be the deadline that the EHD sends out all data. Denote the residual energy at time $t_x\!\in\![0,T]$ by $E_{t_x}^b$, where
\begin{equation}
\label{eq:eh01}
	E_{t_x}^b=\sum_{k=1}^{{t_{r_k}\!<t_x}} E_k^r-	\int_0^{t_x }p^{eh}(t)\,\mathrm{d}t.\\
\end{equation}
According to the charging characteristic of batteries, the amount of energy to be transmitted by an ES depends not only on the amount of energy to replenish but also on the residual energy at the EHD. The relation between $E_i^{es}$, $E_i^r$ and $E_{t_{r_i}}^b$ can be represented by $E_i^{es}\!=\!z(E_i^{r}, E_{t_{r_i}}^b)$, where $z(\cdot)$ is called the charging function and the details will be further elaborated in Section~\ref{sec:PreReq}.

The overall objective of this study is to optimize the energy consumption of the ES under the constraints of data and energy storage capacities at the EHD. This optimization problem, denoted by \textbf{P1}, is formulated as follows:
 \begin{equation}\label{eq:eh02}
\begin{array}{lll}
    \vspace{-0.1cm}
    \textbf{P\,1}:\qquad\underset{t_{r_i}, E_i^r}{\mathrm{arg\,min}}\displaystyle\sum_{i=1}^{t_{r_i}<T}\left(E_i^{es}+e^r\right),\\
    \vspace{-0.1cm}
    \textbf{s.t.}\\
   \vspace{0.05cm}
    \textbf{\emph{C1}:}\;\,\displaystyle\int_0^{t_x}p_o^{eh}(t)\mathrm{d}t\leq\sum_{i=1}^{{t_{r_i}\!<t_x}} E_i^r,\qquad\quad\;\,\, t_x\in[0,T],\\
    \vspace{0.2cm}
    \textbf{\emph{C2}:}\;\;\;\displaystyle E_{t_x}^b\leq E_m^{eh},\qquad\qquad\qquad\qquad\quad\;\,
     t_x\in[0,T],\\
     \vspace{0cm}
    \textbf{\emph{C3}:}\;\;\;T_i^{es}\leq\displaystyle t_{r_{i+1}}\!-t_{r_i},
\end{array}
\end{equation}
where $p_o^{eh}(t)$ is the optimal transmission power that allows an EHD to send all data timely and with minimum energy consumption. Correspondingly, the integration of $p_o^{eh}(t)$ represents the least amount of energy required at the EHD;  $E_m^{eh}$ represents the battery capacity of an EHD and $e^r$ is a constant overhead at the ES to compensate the EHD for the energy spent on transmitting request messages.

In \textbf{P1}, the constraint C1 is that at any time the total energy harvested at the EHD should be no less than the accumulation of the minimum energy it consumed; otherwise the arriving data may overflow the storage. The constraint C2 indicates that the difference between the harvested energy and the expended energy cannot exceed the battery capacity of an EHD, which is referred to as the energy capacity constraint. The constraint C3 is a charging constraint that limits the time interval between two consecutive energy requests. Specifically, as demonstrated in Fig.~\!\ref{fig:EnergyRequest}, the $i^{th}$ energy charging takes $T_i^{es}$ seconds, the EHD thus cannot initiate a new energy request until the current battery replenishment ends.

After a further study, we interpret problem \textbf{P1} as follows: when and how much energy should an EHD request or harvest from the ES, so that the energy consumption at the source is minimized with the three constraints above? To solve problem \textbf{P1}, we propose a two-step energy requesting strategy, DTER, which links the \emph{data transmission} profile at the EHD and the scheduling of \emph{energy transmission} at the ES. Specifically, it is critical to build an analytical relationship between the harvested energy and the energy consumption for data transmissions at the EHD. 
Therefore, the first step of DTER is to calculate the optimal transmission rate at the EHD in order to minimize its cumulative energy consumption for data transmissions.
The profile of optimal transmission rate at the EHD obtained from the first step determines the bounds of an energy tunnel. The second step is to design an optimal energy requesting strategy inside the energy tunnel so that the energy consumption at the ES is minimized as well.

\section{Optimal Transmission Rate}
\label{sec:optTranRate}

In this section, we present the first step of DTER. The objective is to calculate the profile of the optimal transmission rate at the EHD so that its cumulative energy consumption for data transmission is minimized.

\subsection{Feasible Data Tunnel}
\label{subsec:optTranRate_Tunnel}

Assuming the initial energy at the EHD is infinite, we focus on calculating the optimal transmission rate while considering the data capacity constraint. The energy constraint will be integrated into the second step, as presented in Section \ref{sec:EngReq}. We formulate the first step of DTER as an optimization problem, as follows.
\begin{equation}\label{eq:eh03}
\begin{array}{lll}
    \vspace{-0.1cm}
    \textbf{P2:}\qquad\underset{r^{eh}(t)}{\mathrm{arg\,min}}\displaystyle\int_0^T f\left(r^{eh}(t)\right)\,\mathrm{d}t,\\
    \vspace{0.0cm}
    \textbf{s.t.}\\
    \vspace{0.1cm}
    \textbf{\emph{C1}:}\;\,\displaystyle\int_0^{t_x}\!\!r^{eh}(t)\,\mathrm{d}t\,\leq\!\sum_{n=1}^{t_{d_n}\!<t_x}\!\!D_n,\qquad\quad\quad t_x\in[0,T],\\
    \textbf{\emph{C2}:}\;\,\displaystyle\int_0^{t_x}\!\!r^{eh}(t)\,\mathrm{d}t\,\geq\!\sum_{n=1}^{t_{d_n}\!<t_x}\!\!D_n-D_m^{eh},\;\;\;\,t_x\in[0,T].\\
\end{array}
\end{equation}
where $f(\cdot)$ is the inverse function of $g(\cdot)$, i.e., $f(\cdot)\!=\!g^{-1}(\cdot)$, which is non-negative, increasing, continuously differentiable, and strictly convex.

\begin{figure}[htb]
\centerline{\includegraphics[width=6.7cm]{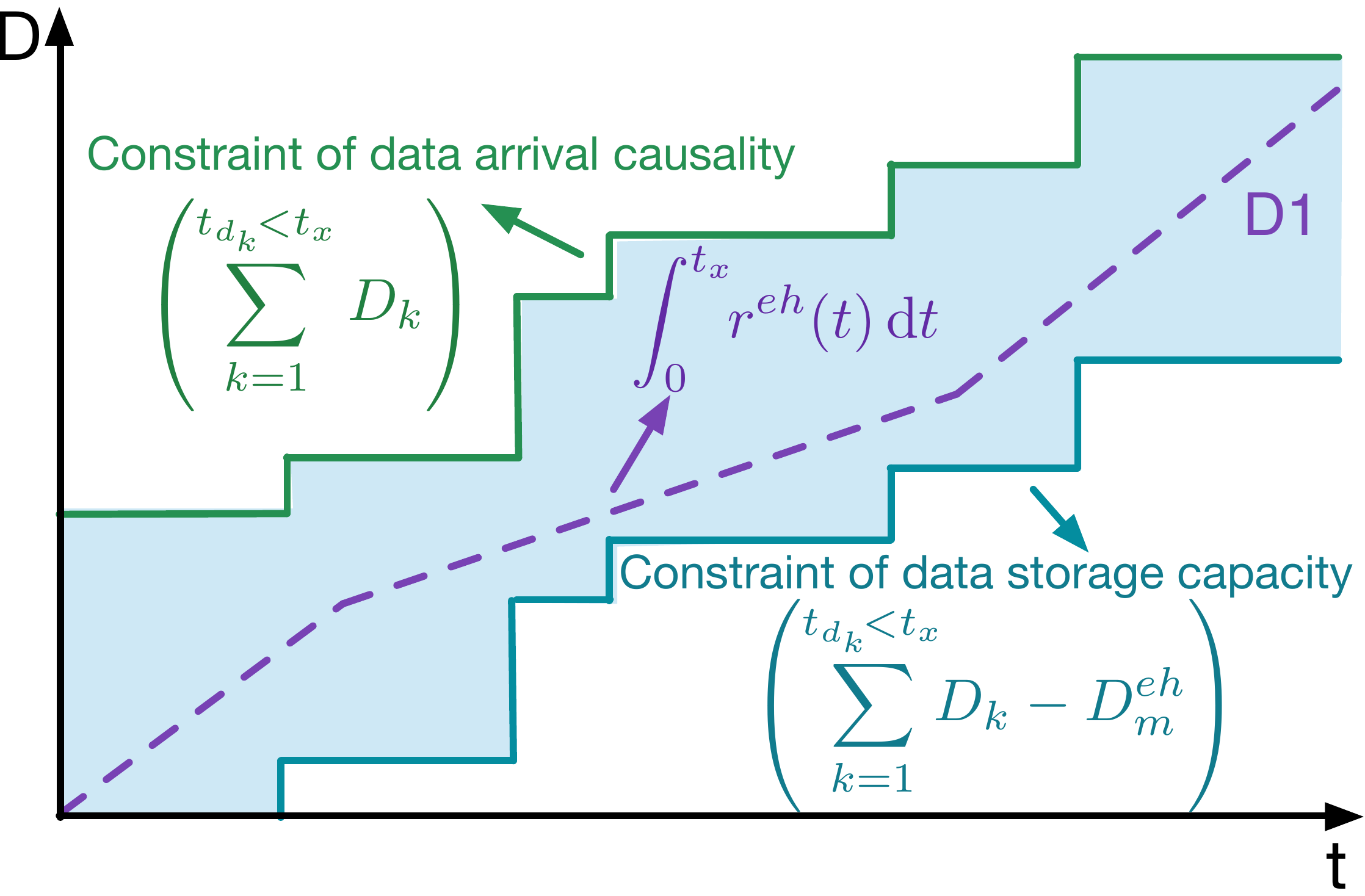}}
\caption{An example of feasible data tunnel.}\label{fig:DataTunnel}
\end{figure}

Here the problem \textbf{P2} is to minimize the energy consumption of the EHD on transmitting all data, subject to two constraints. Constraint C1 reflects the causality of data transmission, i.e., the EHD cannot send data that has not arrived yet. Constraint C2 guarantees that the data arrival does not overflow the data storage. These two constraints construct a feasible data tunnel, as illustrated in Fig.~\!\ref{fig:DataTunnel}.
The profile of cumulative data transmitted at the EHD is a continuous line (e.g., curve D1 in Fig.~\!\ref{fig:DataTunnel}) that stays within the feasible data tunnel to satisfy the data causality and storage capacity constrains.

\subsection{Lemmas of Optimal Transmission Rate}
\label{subsec:optTranRate_Lem}

Here, let $t_p$ be a time point that the sum of data stored in the EHD and all data arriving in the future are equivalent to the maximum capacity of the data storage, i.e.,
\begin{equation}\label{eq:eh04}
	D_{t_p}^b\!+\!\sum_i D_i=D_m^{eh}, \quad\forall i\!\in\!\mathbb{Z}^+,\; t_{d_i}\!\in\!(t_p, T].\end{equation} 
Now, we present five lemmas to explore the inherent features of the optimal transmission rate, $r_o^{th}$, in (\ref{eq:eh03}).
\vspace{0.0cm}
\begin{adjustwidth}{-0cm}{-0cm}
\setlength{\parindent}{0pt}
\begin{lem}
\label{lem:02}
	\emph{The transmission rate is constant during an interval between the arrivals of two successive data packets.}
\end{lem}

\begin{lem}
\label{lem:03}
	\emph{The transmission rate changes only when the data storage is either full or completely depleted.}
\end{lem}

\begin{lem}
\label{lem:04}
	\emph{The transmission rate decreases monotonically if the data storage is not completely depleted, i.e., $\forall\, t_i, t_j\in[0, T], t_i<t_j\!:\, r_o^{eh}(t_i)\geq r_o^{eh}(t_j)$, if  $\forall\, t\in [t_i, t_j]\!: D_t^b\neq 0$.}
\end{lem}

\begin{lem}
\label{lem:05}
	\emph{The transmission rate increases monotonically if the data storage is not completely filled, i.e., $\forall\, t_i, t_j\in[0, T], t_i<t_j\!: r_o^{eh}(t_i)\leq r_o^{eh}(t_j)$, if  $\forall\, t\in [t_i, t_j]\!: D_t^b\neq D_m^{eh}$.}
\end{lem}

\begin{lem}
\label{lem:06}
	\emph{The transmission rate approaches zero after $t_p$.}
\end{lem}
\end{adjustwidth}
\vspace{0.1cm}
Lemma~\!\ref{lem:02} to Lemma~\ref{lem:05} can be proved by the contradiction method; similar proofs can be found in~\cite{yang2012optimal} and \cite{tutuncuoglu2012optimum}. The proof of Lemma~\ref{lem:06} can be found in Appendix~\ref{app:appendix}. From Lemma~\!\ref{lem:02} to Lemma~\ref{lem:05}, the following Corollary is deduced.

\vspace{0.0cm}
\addtocounter{thm}{-4}	
\begin{adjustwidth}{-0cm}{-0cm}
\setlength{\parindent}{0pt}
\begin{corl}
\label{corl:02}
	\emph{At instants of data arrival, the transmission rate decreases if the data storage is completely filled or increases if the storage is empty, i.e., $\forall\, t\in [0, T]\!:\! r_o^{eh}(t^-)\!>\!r_o^{eh}(t^+)\!\!\implies\!\! D_t^b=D_m^{eh}$ and $r_o^{eh}(t^-)\!<\!r_o^{eh}(t^+)\!\!\implies\!\! D_t^b=0$.}
\end{corl}
\end{adjustwidth}
\vspace{0.1cm}

Essentially, the objective of the optimization problem in (\ref{eq:eh03}) is to seek the best curve in a feasible data tunnel to minimize a specified cost, which is similar to the optimization problem solved in \cite{tutuncuoglu2012optimum}. Consequently, even though the two optimization problems have entirely different objective functions and constraints, their results are naturally related from the graphical point of view. This can be observed by comparing the above Corollary \ref{corl:02} with Corollary 1 introduced in \cite{tutuncuoglu2012optimum}. To solve \textbf{P2}, either the water-filling algorithm presented in \cite{ozel2011transmission} or the throughput maximizing method proposed in \cite{tutuncuoglu2012optimum} is a feasible approach. Eventually, it is verified that the profile of the optimal transmission rate, $r_o^{eh}$, is the tightest and piecewise segments that lie in the feasible data tunnel.

\section{Energy Tunnel and Charging Function}
\label{sec:PreReq}

Before presenting the second step of DTER, we introduce two concepts first, namely, \emph{feasible energy tunnel} and \emph{charging function}.
These two concepts are critical to the design of the second step of DTER (presented in the Section VI) to obtain the optimal strategy for energy harvesting.


\subsection{Feasible Energy Tunnel}
\label{subsec:PreReq_Tunnel}

\begin{figure}[htb]
\centerline{\includegraphics[width=6.7cm]{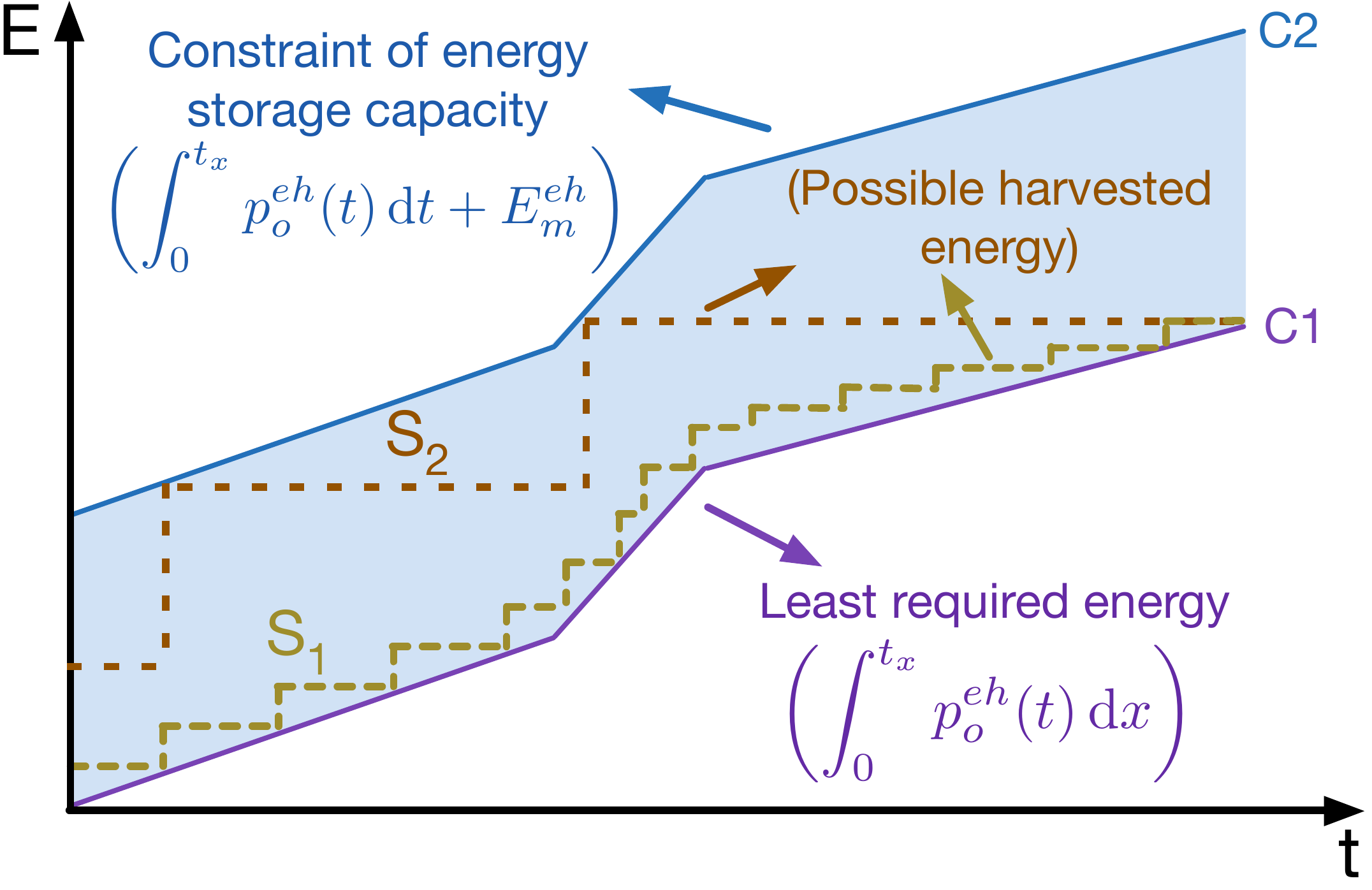}}
\caption{The feasible energy tunnel.}\label{fig:EngTunnel}
\end{figure}

The constraints C1 and C2 in problem \textbf{P1} define the upper bound and lower bound for the accumulation of harvested energy, respectively. The bounded tunnel is termed the \emph{feasible energy tunnel}. As illustrated in Fig.~\!\ref{fig:EngTunnel}, the lower bound is the least required power ($\int_0^{t_x}\!\!p_o^{eh}(t)$), which can be calculated from the optimal transmission rate (i.e., $r_o^{eh}(t)$) derived from the first step of DTER. The upper bound is obtained by shifting the lower bound up by $E_m^{eh}$, indicating the battery capacity constraint. Referring to Section ~\!\ref{subsec:optTranRate_Lem}, the upper bound and lower bound of feasible energy tunnel will be piecewise linear.

The objective in the second step of DTER is to minimize the overall energy consumption of ES, which is achieved by scheduling the stair-stepping profile of harvested energy inside the feasible energy tunnel, as demonstrated in Fig.~\!\ref{fig:EngTunnel}. Finding the optimal requesting strategy is not trivial. 
Specifically, the EHD needs to pay a constant cost on each request message transmission, and consequently generates significant energy overhead if it requests too frequently (e.g., curve S1 in Fig.~\!\ref{fig:EngTunnel}). In contrast, curve S2 in Fig.~\!\ref{fig:EngTunnel} asks too much energy each time resulting in low efficiency on energy transfer as the energy cost at the ES increases nonlinearly with the growth of energy to be charged~\cite{gorlatova2013networking}. Therefore, neither strategy S$_1$ nor S$_2$ is a good option for efficient energy request.


\subsection{Charging Function}
\label{subsec:PreReq_Char}

In RF-based EH systems, an EHD is generally equipped with a super capacitor to store the harvested energy~\cite{mishra2015smart, liu2013ambient}.
Therefore, we use the capacitor as an example to provide insight into the charging function, which determines the efficiency of energy replenishment.

To charge an EHD wirelessly, assume the ES transmits energy packets with a constant power, which is represented as $p^{es}$. The transmission time of the energy packet $i$ is $T_i^{es}$. According to the relationship between the voltage increase and energy replenishment presented as (4) in \cite{luo2017optimal}, the energy cost on an ES for the $i^{th}$ charging is calculated below:
\begin{equation}\label{eq:eh06}
\begin{array}{lll}
    E_i^{es}&\!\!\!\!\!=\!\!\!\!\!&p^{es}T_i^{es}\\
    \vspace{0.15cm}
                  &\!\!\!\!\!=\!\!\!\!\!&\displaystyle p^{es}RC\ln\left\{\frac{\left(2E^{eh}_m\right)^{\frac{1}{2}}\!-\!\left(2E_{t_{r_i}}^b\right)^{\frac{1}{2}}}{\left(2E^{eh}_m\right)^{\frac{1}{2}}\!-\!\left[2\left(E_{t_{r_i}}^b\!+\!E_i^r\right)\right]^{\frac{1}{2}}}\right\},
    \end{array}
\end{equation}
where $R$ and $C$ are the resistance and capacitance of the charging circuit, respectively. $E^{eh}_m\!=\!\frac{1}{2}CV_m^2$. $V_m$ is the maximum voltage a capacitor could approach, which is determined by the EH circuit and the receiving power of energy packets at an EHD~\cite{nintanavongsa2012design}. The above equation verifies that the energy consumption at ES is not only affected by the amount of energy the EHD claimed, $E_i^r$, but also by the residual energy in the capacitor, $E_{t_{r_i}}^b$. 

Consequently, we introduce a useful property of $E_i^{es}$ to support the design of an optimal energy requesting strategy.
\vspace{0.1cm}
\begin{adjustwidth}{-0cm}{-0cm}
\setlength{\parindent}{0pt}
\addtocounter{thm}{-1}    
\begin{pro}
\label{pro:01}
    \emph{$E_i^{es}$ is an increasing function of $E_i^r$ and a strictly convex function of $E_{t_{r_i}}^b$. With regard to $E_i^r$, the $E_i^{es}$ function is strictly concave if $E^r_i+E_{t_{r_i}}^b\!\in[0,\frac{1}{4}E_m^{eh})$, and strictly convex if $E_i^r+E_{t_{r_i}}^b\!\in[\frac{1}{4}E_m^{eh}, E_m^{eh}]$.}
\end{pro}
\end{adjustwidth}
\vspace{0.1cm}
The verification of Property~\ref{pro:01} is straightforward. We can prove it by inspecting the first order and the second order partial derivatives of $E_i^{es}$ in (\ref{eq:eh06}) with respect to $E_i^r$ and $E_{t_{r_i}}^b$.

Property~\ref{pro:01} and (\ref{eq:eh06}) indicate that with  fixed lengths of energy packets, the energy harvested by an EHD does not decrease monotonically with the increase of $E_{t_{r_i}}^b$. In other words, to improve the efficiency of energy harvest, the EHD needs to keep a certain amount of energy in its capacitor.

\section{Energy Requesting Strategy}
\label{sec:EngReq}

The second step of DTER focuses on how to design the optimal strategy for energy requesting. For that, we first investigate the piecewise optimal strategies without the overhead energy, which provides an upper bound of the data transmission rate that a capacitor-based EHD can achieve. Then we propose a heuristic approach that converts the scheduling of energy requests into a shortest path problem, in which a dynamic programming (DP) algorithm is applied to achieve the global optimal solution.

\subsection{Piecewise Optimal Strategy with Negligible Overhead}
\label{subsec:EngReq_LarCos}

This section considers a simplified scenario where the overhead energy, $e^r$, is assumed negligible. We aim to minimize the energy consumption at the ES in each piece of the feasible energy tunnel separately. Since the slope for the same piece is fixed, $E^b_{t_{r_i}}$ and $E_i^r$ will remain unchanged and their subscripts can be dropped.

According to Property~\!\ref{pro:01}, in the case that the constant overhead, $e^r$, is negligible, an EHD harvests energy most efficiently when it requests a small amount of energy if and only if a quarter of maximum energy remains in the capacitor. Correspondingly, the following corollary is deduced.

\addtocounter{thm}{0}    
\vspace{0.1cm}
\begin{corl}
\label{corl:03}
    \emph{For an optimal energy requesting strategy with negligible $e^r$, the EHD requests a tiny amount of energy from ES each time, when the residual energy is $\frac{1}{4}E_m^{eh}$, i.e, $E^r \!\to\! 0$ and $E^b=\frac{1}{4}E_m^{eh}$, if $e^r\to 0$.}
\end{corl}
\vspace{0.1cm}
Corollary \ref{corl:03} forms the foundation to obtain the upper bound of a transmission rate for a capacitor-based EHD, which will be derived as follows.
From Corollary \ref{corl:03}, the EHD cannot consume more energy during an energy request interval than the amount harvested in the last round, which is $E^r$. Combining this charging constraint with the constraint of the least energy requirement at the EHD, we have
\begin{equation}\label{eq:eh07}
	\displaystyle \frac{E^r}{T^{es}\,p^{eh}}\geq1,
\end{equation}
where
\begin{equation}\label{eq:eh08}
    T^{es}=\displaystyle RC\ln\left\{\frac{\left(2E^{eh}_m\right)^{\frac{1}{2}}-\left(2E^b\right)^{\frac{1}{2}}}{\left(2E^{eh}_m\right)^{\frac{1}{2}}-\left[2\left(E^b+E^r\right)\right]^{\frac{1}{2}}}\right\},
\end{equation}
which is calculated from (\ref{eq:eh06}).

According to Corollary~\ref{corl:03}, substituting $E^b\!=\!\frac{1}{4}E_m^{eh}$ and $E^r \to 0$ into $T^{es}$ of  (\ref{eq:eh07}), the left side of the inequalities is a limit of the form $0/0$. Utilizing the continuously differentiable feature of $T^{es}$ with respect to $E^r$ and applying the L'H\^opital's rule on the left side of (\ref{eq:eh07}), we derive
\begin{equation}\label{eq:eh09}
	\displaystyle\lim_{E^r\to0} \frac{E^r}{T^{es}\,p^{eh}}=\displaystyle\lim_{E^r\to0}\frac{1}{(T^{es})^\prime\,p^{eh}}=\displaystyle \frac{E_m^{eh}}{2\,R\,C\,p^{eh}},
\end{equation}
where $(T^{es})^\prime$ is the first order derivation of $T^{es}$ with respect to $E^r$. Substituting (\ref{eq:eh09}) into (\ref{eq:eh07}), we obtain
\begin{equation}\label{eq:eh10}
	p^{eh}_m(t)\leq\frac{E_m^{eh}}{2\,R\,C}=\frac{V^2_m}{4R}.
\end{equation}
In (\ref{eq:eh10}), $p^{eh}_m(t)$ is the highest power that a capacitor-based EHD could harvest from a dedicated ES. It reveals the upper bound of the energy consumption in a given time period regardless of the requesting strategy. Substituting (\ref{eq:eh10}) into the power-rate function, $g(\cdot)$, it yields an upper bound of the data transmission rate that a capacitor-based EHD can achieve.

\subsection{Global Optimal Strategy}
\label{subsec:EngReq_Opt}
Due to the discrete feature of the harvested energy and the nonlinear relationship between the energy transmitted from an ES and that replenished to an EHD, the design of a global optimal strategy is a grand challenge. To tackle this challenge, we leverage the Graph theory to convert the optimal energy requesting design into a shortest path problem.

\begin{figure}[htb]
\centerline{\includegraphics[width=6.7cm]{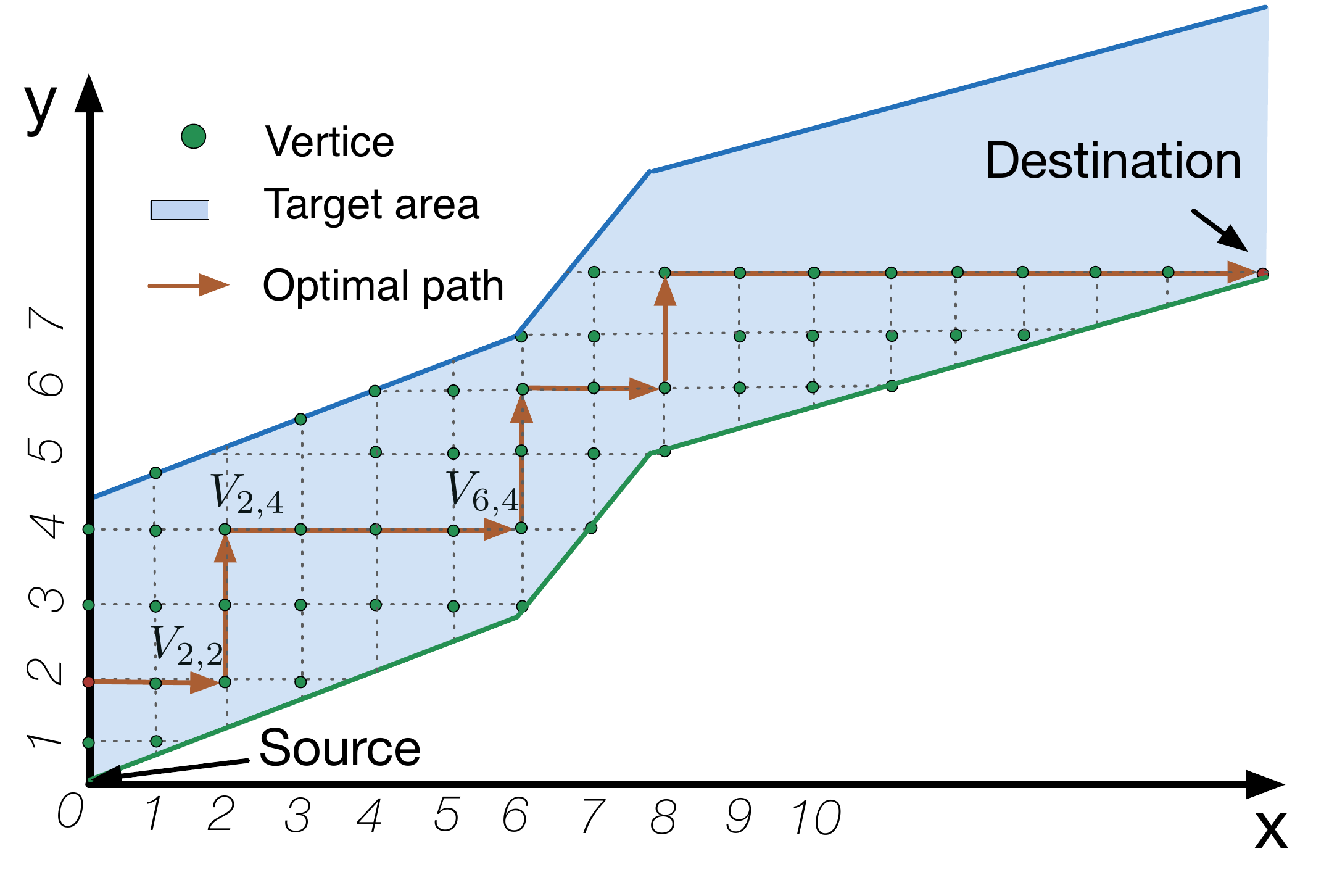}}
\caption{The converted shortest path problem in a feasible energy tunnel.}\label{fig:StrategyThree}
\end{figure}

As depicted in Fig.~\ref{fig:StrategyThree}, assume the slope of each energy tunnel is smaller than the maximum transmission power in (\ref{eq:eh10}) that an EHD can afford, and then the area of the energy tunnel is divided into multiple grids uniformly. The intersections of the grids are referred to as ``vertices'' connected with horizontal and vertical ``edges''. The amount of grids depends on the requirement of the solution accuracy and will be evaluated from simulations. Denote a vertex with the coordinates $(j, k)$ by $V_{j, k}$, and the directed edge from $V_{j, k}$ to $V_{m, n}$ is represented by $e_{(j, k)\to(m, n)}$, where $\forall j, k, m, n\in \mathbb{Z}^+$.  A vertical edge $e_{(j, k)\to(j, n)}$ indicates an energy replenish by $E_i^r=y_n-y_k$, which generates an associate charging cost on an ES. We call this cost the weight of the edge, the calculation of which is specified by (\ref{eq:eh06}). A horizontal edge means no charging and has a weight of $0$. Therefore, the weight of  $e_{(j, k)\to(m, n)}$ is
\begin{equation}\label{eq:eh11}
w_{(j, k)\to(m, n)}\!=\!
    \left\{\!\!
    \begin{array}{lll}
        \vspace{0.1cm}
        0, \qquad\qquad\quad &j&\!\!\!\!\neq m\, \,\text{and}\; k=n,\\
        E_i^{es}, \qquad\quad  &k&\!\!\!\!\neq n\, \,\text{and}\;j=m.
    \end{array}
    \right.
\end{equation}

A node is placed at the source with coordinates $(0, 0)$ and then moves towards the destination at the right end of the energy tunnel's lower boundary. It is not allowed to move backwards or downwards according to the features of EH process. Due to the charging constraint introduced in C3 of \textbf{P1}, the length of a lateral movement cannot be shorter than the charge time of the last energy replenishment, i.e., $T_i^{es}$.

So far the second step of DTER has been converted into the classic shortest path problem in a directed and weighted graph. Designing the optimal strategy for energy requesting is equivalent to finding the shortest path from the source to the destination such that the sum-weight of its constituent edges is minimized under the charging constraint. Consequently, the DP-based approach like Dijkstra's algorithm can be applied to schedule the optimal energy request scheme. The time complexity of such approach is $\mathcal{O}(n\log{n})$, where $n$ is the number of grids in the energy tunnel. More details of DP algorithms could be found in \cite{sniedovich2010dynamic}.

\section{Online Suboptimal Strategy}
\label{sec:OnlineOpt}

The global optimal strategy for the second step of DTER presented in Section~\ref{subsec:EngReq_Opt} provides a lower bound of the energy consumption that an EH system can achieve. It is an offline strategy since the future transmission power of an EHD is assumed known as a priori knowledge, which is impractical for some applications. In this section,
we propose an online suboptimal strategy, which does not require any future information of EH system and just needs to know the current energy level in the capacitor of an EHD.

Here, we start from a common scenario where the overhead energy for each energy request is not negligible with the goal of minimizing the total energy consumption at the ES in each piece of energy tunnel separately. Denote the number of requests that an EHD initiates in the $k^{th}$ piece of tunnel by  $\alpha_k$. Assume the $k^{th}$ piece of feasible energy tunnel starts from $t_{p_{k-1}}$ and ends at $t_{p_k}$. Let $p_k^{eh}$ and $l_k$ represent the slope (e.g., optimal transmission power of an EHD) and duration of the tunnel piece $k$, where $l_k=t_{p_k}-t_{p_{k-1}}$, we obtain that 
    \begin{equation}\label{eq:eh12}
        \alpha_k=\displaystyle\frac{l_k\,p_k^{eh}}{E_i^r},\quad \forall i \in \mathbb{Z}^+\!\!: t_{r_i} \!\in\!\left(t_{p_{k-1}},\, t_{p_k}\right].
    \end{equation}

As in the scenario with negligible overhead that discussed in Section~\ref{subsec:EngReq_LarCos}, for a certain piece of feasible energy tunnel, the subscripts of $E^b_{t_{r_i}}$ and $E_i^r$ can be removed. In other words, the optimal $E^b_{t_{r_i}}$ and $E_i^r$ are constants in a single energy tunnel. Based on (\ref{eq:eh06}), the optimization problem \textbf{P1} is rewritten as
 \begin{equation}\label{eq:eh13}
\begin{array}{lll}
    \textbf{P\,3}\!:
    \vspace{0.3cm}
    \underset{E^r,\,E^b}{\mathrm{arg\,min}}\,\,\alpha_k\!\!\left[ p^{es}\!RC\displaystyle\ln\!\left\{\!\!\frac{\left(2E^{eh}_m\right)^{\frac{1}{2}}\!-\!\left(2E^b\right)^{\frac{1}{2}}}{\left(2E^{eh}_m\right)^{\frac{1}{2}}\!\!-\!\!\left[2\!\left(E^b\!+\!E^r\right)\right]^{\frac{1}{2}}}\!\!\right\}\!+e^r\!\right],\\
    \begin{array}{lll}
    \vspace{0.2cm}
    \hspace{-0.1cm}
    \textbf{s.t.}\quad
    &\textbf{\emph{C1}:}\quad E_m^{eh}\geq E^b+E^r,\\
    \vspace{0.1cm}
    &\textbf{\emph{C2}:}\quad E^b\geq0, \;E^r\geq 0 \;\;\text{and}\;\; \alpha_k \in\mathbb{Z}^+,\\

    &\textbf{\emph{C3}:}\quad \displaystyle E^r\geq T^{es}\,p_k^{eh}.
    \end{array}
    \end{array}
\end{equation}
In problem \textbf{P3}, C1 and C2 correspond to the constraints of data capacity and least energy requirement, respectively. The constraint C3 represents the charging constraint, which is equivalent to C3 of \textbf{P1}.

The objective function in \textbf{P3} has a global minimum since its Hessian matrix is positive definite with respect to $E^b$ and $E^r$. To calculate the optimal $E^b$ and $E^r$, we set the first order partial derivatives of the objective function to zero and obtain
\begin{equation}\label{eq:eh14}
    \begin{cases}
    \vspace{0.15cm}
    \displaystyle E^b=\displaystyle\frac{\left(E_m^{eh}-E^r\right)^2}{4E_m^{eh}}, &\mbox{ }\\

    \displaystyle (\ln X)-\frac{X^2-1}{2X}+\frac{e^r}{R\,C\,p^{es}} =0, &\mbox{ }
    \end{cases}
\end{equation}
where
\begin{equation}\label{eq:eh15}
   E^r=\displaystyle\frac{(X-1)}{(X+1)}\,E_m^{eh}.
\end{equation}
The second equation of (\ref{eq:eh14}) is nonlinear and the explicit solution may not be available. Newton's iterative method is utilized to reach a numerical result, which will be substituted into (\ref{eq:eh15}) to calculate $E^r$, and the result is denoted by $E^r_x$. Note that $E^r_x$ is calculated solely based on the objective function of \textbf{P3} without considering the constraints. Next, we calculate an $E^r_y$ merely based on the constraints.
By combining $E^r_x$ and $E^r_y$, the optimal $E^r$ and $E^b$ will eventually be achieved.

Combining (\ref{eq:eh06}) and the first equation of (\ref{eq:eh14}), the constraint C3 of \textbf{P3} can be rewritten as
\begin{equation}\label{eq:eh16}
	p_k^{eh}RC\ln\left(\frac{E_m^{eh}+E^r}{E_m^{eh}-E^r}\right)-E^r\leq0.
\end{equation}
Let $z(E^r)$ be the left part of (\ref{eq:eh16}) and its first order derivative with respect to $E^r$ is
\begin{equation}\label{eq:eh17}
   z^\prime(E^r)=\displaystyle\frac{E_m^{eh}\left(2p_k^{eh}RC-E_m^{eh}\right)+(E^r)^2}{\left[(E_m^{eh})^2-(E^r)^2\right]}.
\end{equation}
It can be observed that $z(E^r)$ is a strictly convex function of $E^r$ that $z(0)=0$ and $z(E_m^{eh})\to\infty$. In addition, according to (\ref{eq:eh10}), we have that $-E_m^{eh}<2p_i^{eh}RC-E_m^{eh}<0$, i.e., $z^\prime(E^r)$ changes from negative to positive with the increase of $E_m^{eh}$ since $E^r\!\leq\!E_m^{eh}$. Therefore, $z^\prime(E^r)$ is a non-monotonic strictly convex function, and there exists one and only one $E_y^r\in(0, E_m^{eh})$ that makes $z(E_y^r)=0$. The value of $E_y^r$ can be calculated via Newton's iteration. According to (\ref{eq:eh12}), the optimal $\alpha_k$, which is denoted by $\hat{\alpha}_k$ is
\begin{equation}\label{eq:eh18}
	\hat{\alpha}_k=\displaystyle\frac{l_k\,p_k^{eh}}{\min\{E^r_x,\,E^r_y\}}.
\end{equation}
Finally, the optimal $E^r$, which is denoted by $\hat{E}^r$ is $\hat{E^r}\!=\!\min\{E^r_x,\,E^r_y\}$, while the optimal $E^b$, represented by $\hat{E}^b$, is calculated from the first equation of (\ref{eq:eh14}).

Note that if the current transmission power of an EHD is much smaller than the instant charging rate of capacitor, i.e., $p^{eh}_k\!\ll\!p^c_i$, $\forall i \!\in\! \mathbb{Z}^+\!\!: t_{r_i} \!\in\!\left(t_{p_{k-1}},\, t_{p_k}\right]$, the charging constraint can be satisfied inherently and hence $\hat{E}^r\!=\!E^r_x$. In this case, $\hat{E}^r$ and $\hat{E}^b$ are affected neither by the length nor by the slope of the tunnel, since the second equation in (\ref{eq:eh14}) does not contain the parameters $l_k$ and $p_k^{eh}$. This indicates that the piecewise optimal energy request discussed in this section essentially becomes an online strategy. Specifically, the EHD only needs to monitor the residual energy and requests $\hat{E}^r$ energy from ES when the energy left in the capacitor drops to $\hat{E}^b$. 

\begin{figure}[htb]
\centerline{\includegraphics[width=6.7cm,height=4.1cm]{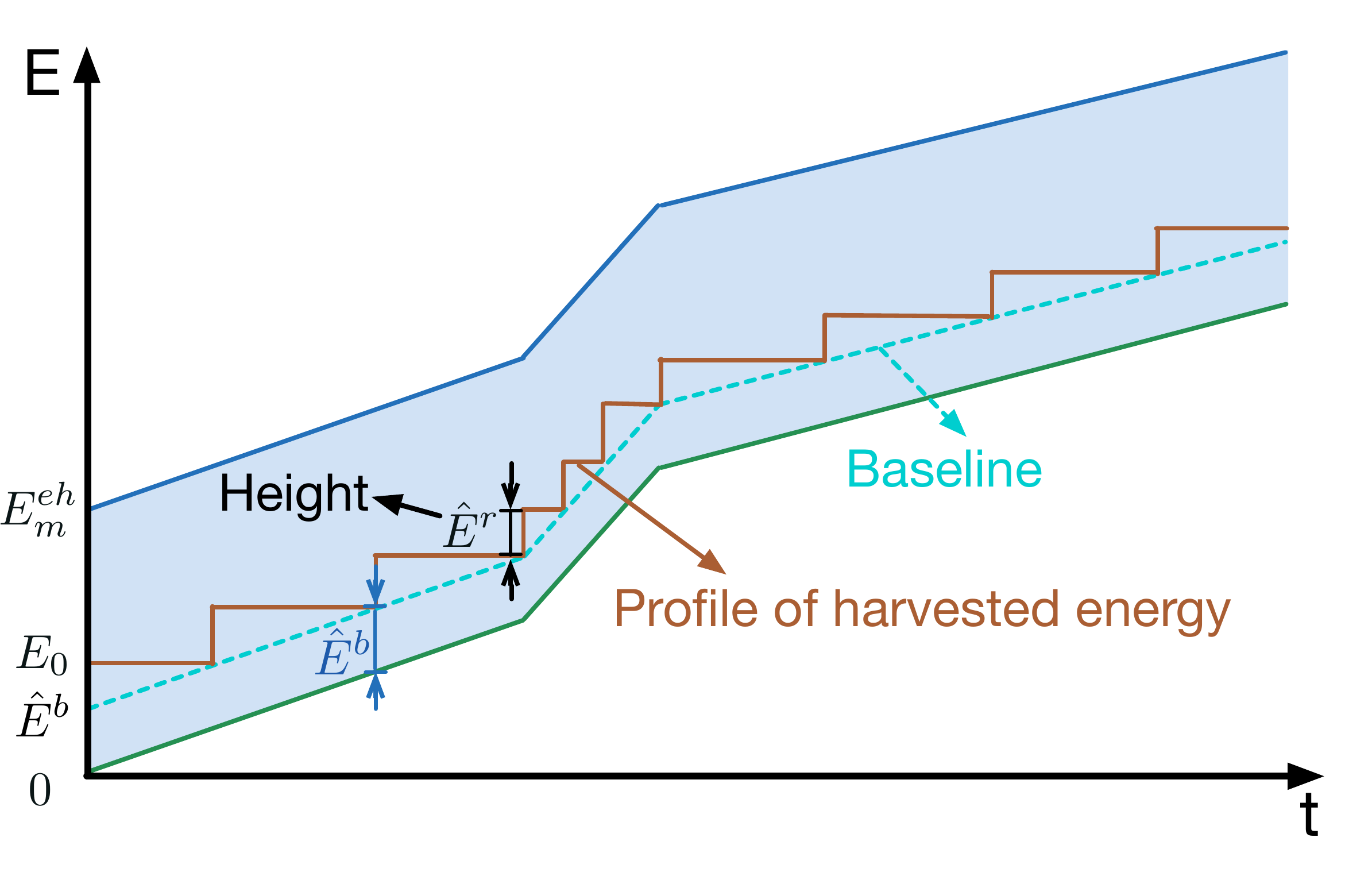}}
\caption{Schematic diagram of the suboptimal online strategy, where the staircase is the profile of cumulative energy harvested by an EHD.}\label{fig:StrategyTwo}
\end{figure}

To facilitate understanding, Fig.~\!\ref{fig:StrategyTwo} illustrates several important attributes of the online energy requesting strategy. As we have discussed in this section, the profile of cumulative energy received by an EHD is a staircase in the figure that has two key features: (a) the amount of energy harvested by an EHD (i.e.,  $\hat{E}^r$) is a constant value from all requests; and (b) the residual energy is also a constant (i.e., $\hat{E}^b$), at the instance of each energy request.

\section{Performance Assessment}
\label{sec:PerEval}

\begin{figure*}[htp]
  \begin{minipage}[t]{0.5\linewidth}
    \centering
    \includegraphics[width=6.4cm, height=5.2cm]{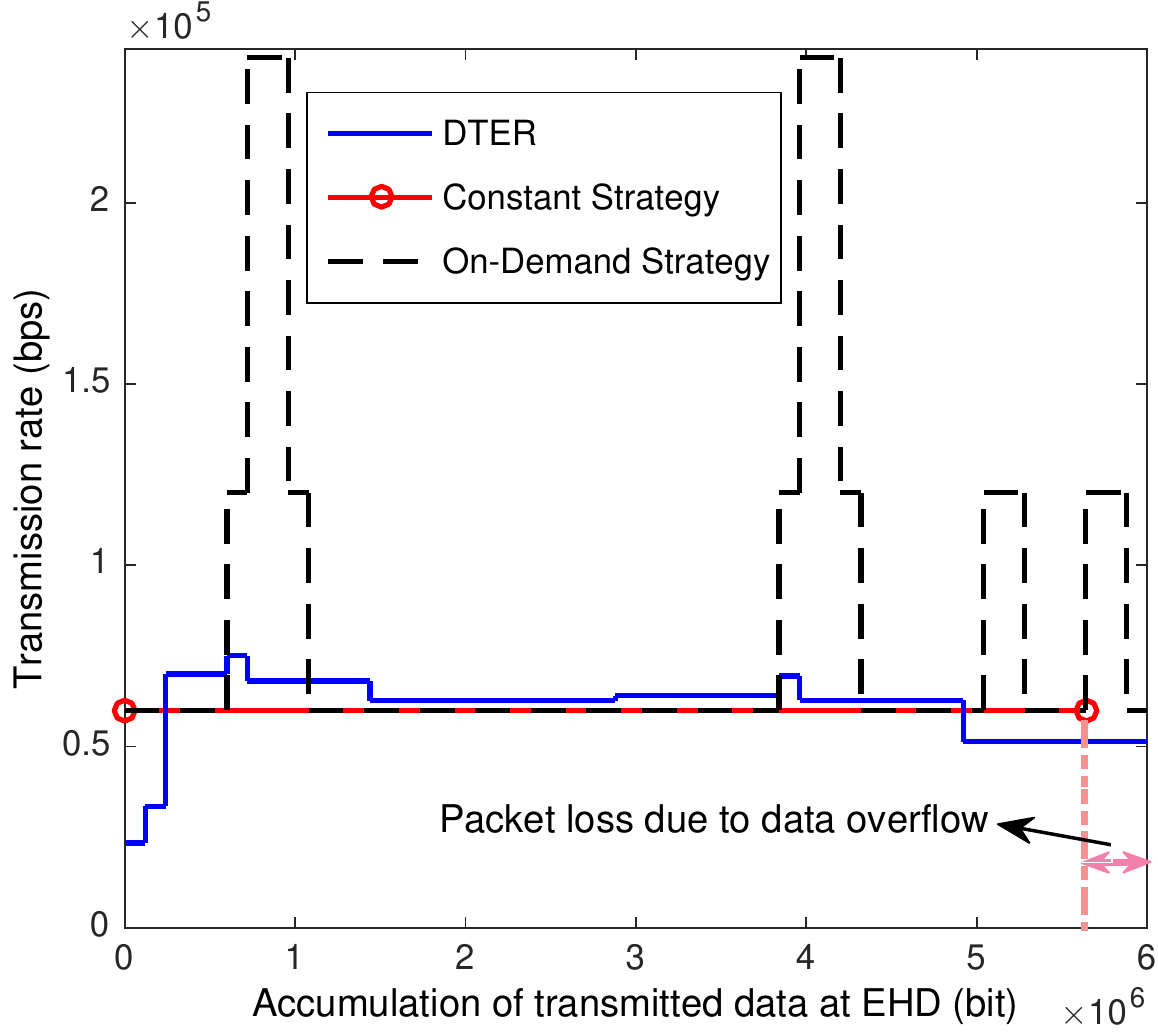}
    \caption{Comparison of transmission rate in a sample run.}
    \label{fig:PerComp:b}
  \end{minipage}%
  \begin{minipage}[t]{0.5\linewidth}
    \centering
    \includegraphics[width=6.4cm, height=5.2cm]{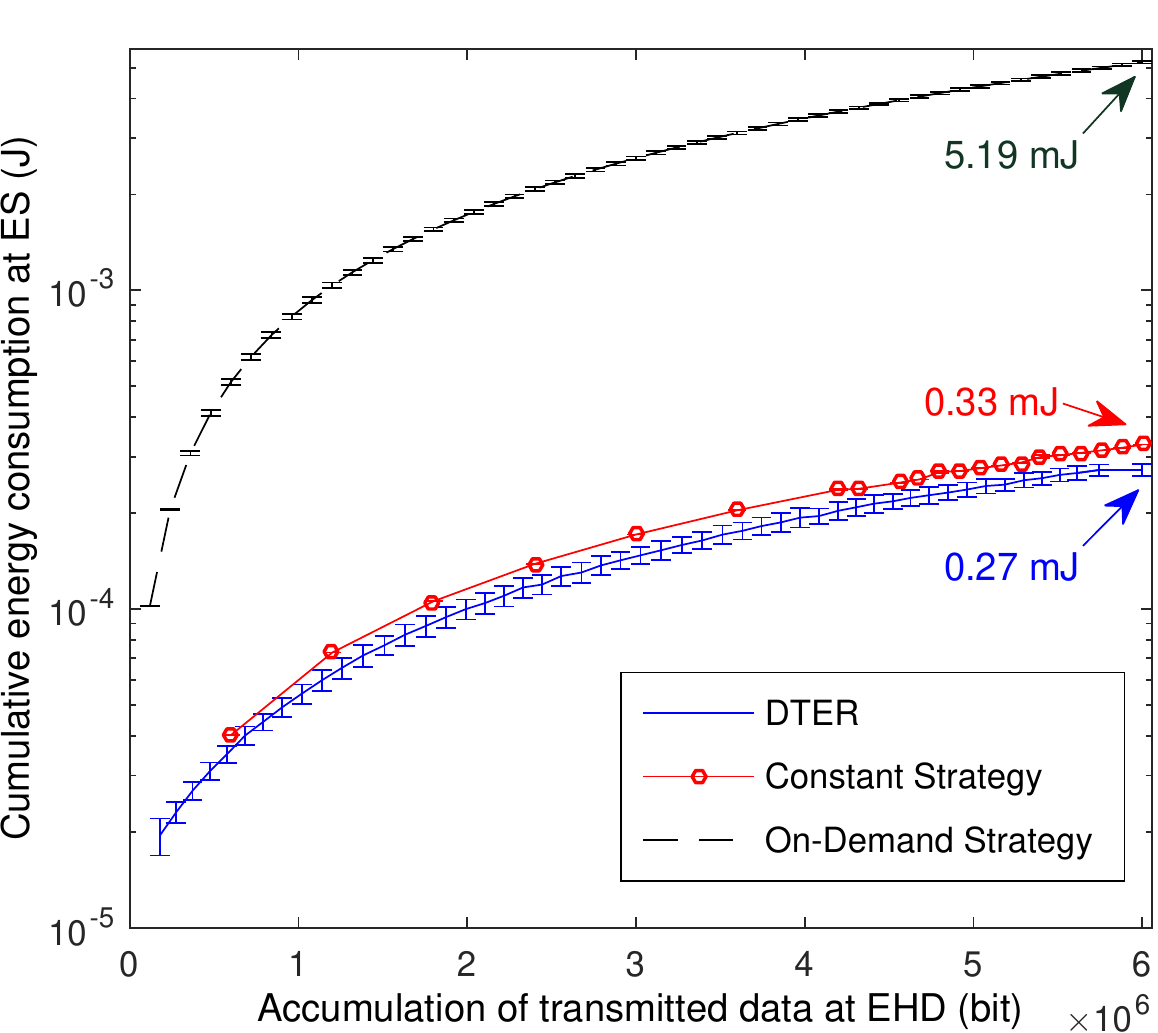}
    \caption{Overall energy consumption among different strategies.}
    \label{fig:PerComp:a}
  \end{minipage}
\end{figure*}

In this section, we conduct simulations to evaluate the performance of DTER. The simulation settings will be introduced first, and then the performance comparisons among different energy requesting strategies are presented. Afterward, we study how different parameters affect the grid density of DTER to achieve the desired accuracy. Lastly, the energy efficiency of the online strategy developed for the second step of DTER is assessed carefully.

\subsection{Simulation Settings}
\label{subsec:PerEval_Setting}

In the simulations, the central frequency and the bandwidth for data transmission are $f_d\!=\!2.4$\,GHz and $B\!=\!50$\,kHz. Assume the wireless channel is AWGN and the noise spectrum density is $N_0\!=\!-174$\,dBm; therefore, the noise power is $N_l\!=\!-127$\,dBm\footnote{$N_l=N_0+10\log_{10} (B)=-127$\,dBm.}. The distance from the EHD to its intended receiver is $d\!=\!30$\,ft, and the propagation loss of an RF signal from EHD to the receiver is calculated through the free space path loss (FSPL) model, where
\begin{equation}\label{eq:eh19}
    \text{FSPL\,(dB)}=20\log_{10} (d)+20\log_{10} (f_d)-147.55.
\end{equation}
According to the capacity of AWGN channel and propagation loss model, the power-rate function is
\begin{equation}\label{eq:eh20}
    p^{eh}\text{(dBm)}\!=\!\displaystyle10\log_{10}\!\left(\!2^{\frac{r^{eh}}{B}}\!\!-\!1\!\right)\!+\text{FSPL}\!+\!N_l.
\end{equation}

We assume the arrival of data packet is a Poisson process with a mean value $\lambda$. The size of each packet and the capacity of data storage are  $S\!=\!15$\,KB and $D_m^{eh}\!=\!64$\,KB. The maximum voltage, capacitance, and resistance of the EHD are $V_m\!=\!2$\,V, $C\!=\!2$\,nF, and $R\!=\!1$\,k$\Omega$, respectively; therefore, the capacitor can store up to $E_m^{eh}\!=\!4$\,nJ of energy. The default overhead, $e^r$, of energy request is $0.4$\,nJ. The ES sends energy packets at a constant power\footnote{$p^{es}$ is determined by the input power and the output voltage of the EHD as introduced in\cite{nintanavongsa2012design}. The output voltage is $2$\,V if the receiving power is around $-10$\,dBm. Therefore, after considering the propagation loss, the required transmission power of the ES is around $40$\,dBm, i.e., $10$\,W.} of $p^{es}\!=\!10$\,W. The simulation results presented in this paper are based on the average of $70$ independent runs, unless otherwise stated.

For comparison purposes, the performance of the following  representative power management strategies are also tested.
\begin{adjustwidth}{-0.18cm}{0cm}
    \begin{itemize}
    \vspace{0.12cm}
    \item[\textendash] \emph{Constant strategy:} In this strategy, the EHD transmits at a constant rate, which is $\lambda S$. When the residual energy is not enough for a single packet transmission, the EHD will request a replenishment and charge the capacitor to $75\%$ of the maximum capacity.
    \vspace{0.12cm}
    \item[\textendash] \emph{On-demand dynamic strategy:} To avoid the overflow of arriving data when the traffic rate has a sudden increase, the EHD adjusts its transmission rate adaptively based on the status of data storage --- the higher occupancy of the storage, the higher transmission rate is applied. The occupancies of data storage and the corresponding transmission rates are $D^b\!=\!\left[\sfrac{1}{16}, \sfrac{1}{8}, \sfrac{1}{4}, \sfrac{1}{2}, \sfrac{3}{4}, \sfrac{7}{8}, \sfrac{15}{16}\right]$ and $r^{eh}\!=\!\left[\sfrac{1}{8}, \sfrac{1}{4}, \sfrac{1}{2}, 1, 2, 4, 8\right]\!\times\!\lambda S$. With this strategy, the EHD is scheduled to request adequate energy for the next data transmission.
 \end{itemize}
 \end{adjustwidth}

\subsection{Performance Comparison}
\label{subsec:PerEval_Comp}

In the performance comparison, we set the traffic generation rate as $\lambda\!=\!0.5$\,pkt/s, and then compare the DTER with two alternative strategies in terms of transmission rate, packet loss ratio caused by the data overflow on the EHD, and energy consumption at the ES.

Fig.~\!\ref{fig:PerComp:b} demonstrates the transmission rate of the EHD under three strategies. According to Corollary \ref{corl:02} in Section~\ref{sec:optTranRate}, low transmission rate and small rate variation are helpful to reduce the energy consumed on data transmission. The constant strategy has the most stable transmission rate, thereby reducing the energy consumption of the EHD considerably. However, it has a significant data overflow problem since the arrivals of data packets are random but the constant strategy cannot adapt to the traffic dynamics. This can be observed in Fig.~\!\ref{fig:PerComp:b} where the accumulation of transmitted data with the constant strategy is less than the other two strategies. According to the measurement, due to the overflow of data storage, the average packet loss ratio of the constant strategy reaches $12.2\%$. In order to send packets timely, the on-demand strategy transmits at varying data rates depending on the occupancy rate of data storage. This causes frequent and drastic changes in the transmission rate at the EHD since there are large-scale fluctuations in the instantaneous arriving rate of data packets. On the contrary, the optimal transmission rate in DTER can adapt to the dynamic traffic while minimizing the variations in transmission rate without overflowing the data storage.

\begin{figure*}[htp]
\centering
\subfigure [Different traffic generation rates.]{
\label{fig:CdfDensity_lambda}
\includegraphics[width=5.8cm]{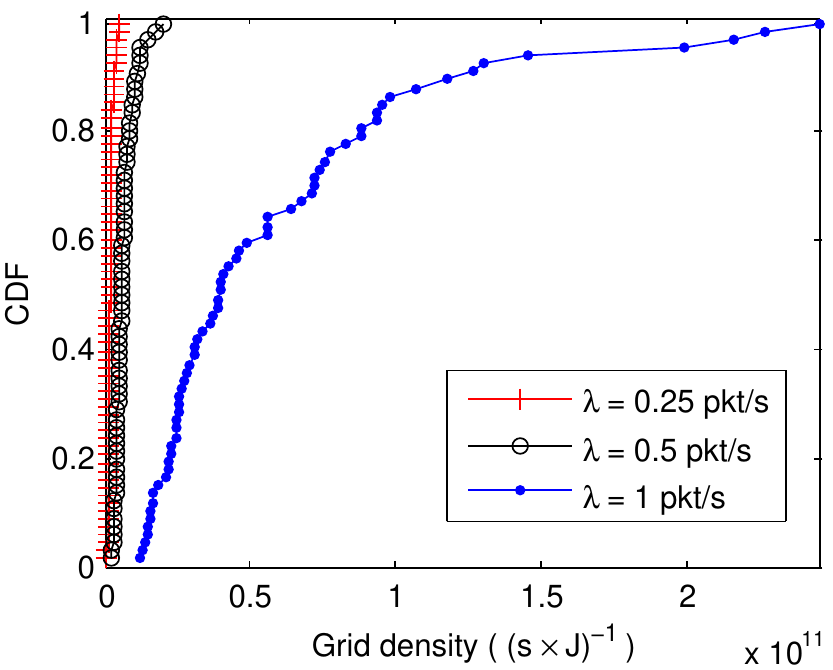}
}%
 \hspace{-0.05in}
\subfigure [Different capacitances.]{
\label{fig:CdfDensity_C}
\includegraphics[width=5.8cm]{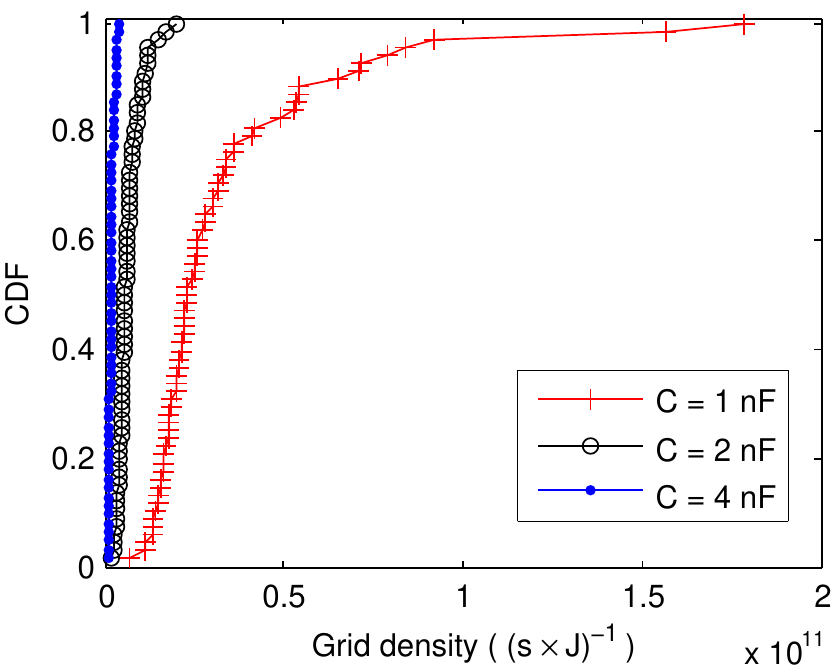}
}%
 \hspace{-0.05in}
\subfigure [Different deadlines.]{
\label{fig:CdfDensity_T}
\includegraphics[width=5.8cm]{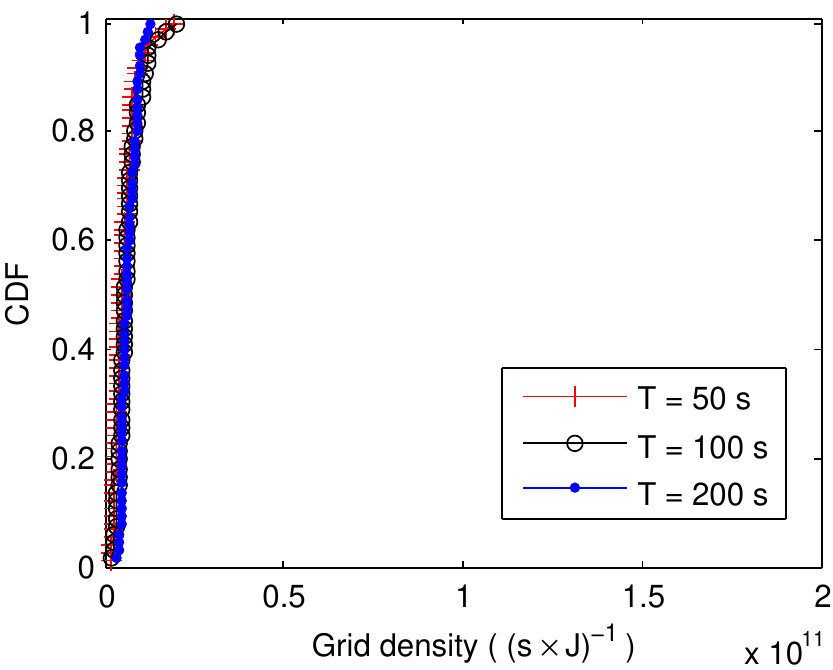}
}%
\caption{CDF of grid density with different settings. (a) $C=2$\,nF, $T=100$\,s; (b) $\lambda=0.5$\,pkt/s, $T=100$\,s; (c) $C=2$\,nF, $\lambda=0.5$\,pkt/s.}
\label{fig:CdfDensity}
\end{figure*}

In Fig.~\!\ref{fig:PerComp:a}, we compare the energy consumption amongst different strategies.
Note that the logarithmic scale is employed for the y-axis due to the huge difference of absolute values. As shown in the figure, DTER has the least energy consumption among the three strategies, mainly for two reasons. First of all, DTER has the optimal transmission strategy at the EHD which guarantees the minimum energy consumption to send a given amount of data. Second, DTER implements the optimal energy requesting strategy which further reduces the energy consumption on ES. Although the results between the DTER and the constant strategy appear to be close to each other in the logarithmic scaled plot in Fig.~\!\ref{fig:PerComp:a}, the actual energy consumption of the constant strategy is much higher than that of DTER, even at the cost of packet loss. More specifically, as shown in the figure, the average energy consumed at ES with the constant strategy is $0.33$\,mJ by the end of tests, $22\%$ higher than that with DTER, which is $0.27$\,mJ. In contrast, the on-demand strategy mitigates the data overflow problem of the constant strategy, but consumes highest energy, which is $5.19$\,mJ at the ES due to the strong fluctuations in transmission rate and inefficient energy requests.

 \subsection{Analysis on Grid Density in DTER}
\label{subsec:PerEval_Density}

In the simulations, Dijkstra's DP algorithm is applied to find the global optimal solution. When using such an approach in DTER, the computing accuracy, i.e., how close is the obtained result to the truly optimal solution, depends on the density of grids, which is defined as the number of grids in the unit area. The result converges to the global optimum energy consumption with the increase of grid density; an excessively high grid density, however, significantly increases the time complexity of the DP algorithm, but gives little improvement on the accuracy. Next, we study how different parameters affect the required density of the gird in DTER to reach a desired accuracy.

Obviously, in order to get a close result to the optimal solution, the length of vertical edge has to be smaller than the optimal $E^r$ and the length of horizontal edge should be smaller than the time interval between neighboring energy requests in the optimal energy requesting strategy. Therefore, the parameters, e.g., $\lambda$ and $C$, that affect $E^r$ and the frequency of energy requests will also affect the required grid density.

As shown in Table.~\!\ref{tab:FreqAmount}, the increased $\lambda$ results in a higher transmission rate and steeper feasible energy tunnel, which in turn increases the frequency of energy requests but does not significantly affect $E^r$. By contrast, a larger $C$ will increase the energy replenished in each round and reduce the energy request frequency accordingly. Therefore, to achieve a certain accuracy, the required density of grid in DTER is proportional to $\lambda$ but inversely proportional to $C$. This conclusion will be verified by the following simulations.

\begin{table}[htp]
\footnotesize
\centering
\caption{Frequency and amount of energy request}
\label{tab:FreqAmount}
\begin{tabular}{|
>{\columncolor[HTML]{D3E8EE}}c |c|c||
>{\columncolor[HTML]{D3E8EE}}c |c|c|}
\hline
\multicolumn{3}{|c||}{\cellcolor[HTML]{D9E7D9}$C=2$\,nF} & \multicolumn{3}{c|}{\cellcolor[HTML]{D9E7D9}$\lambda=0.5$\,pkt\,/\,\!s} \\ \hline\hline
\cellcolor[HTML]{F0DDDC}$\lambda$ & \cellcolor[HTML]{F0DDDC}\begin{tabular}[c]{@{}c@{}}$\hat{E}^r$ (\,$10^{-10}$\,J\,)\end{tabular} & \cellcolor[HTML]{F0DDDC}Freq & \cellcolor[HTML]{F0DDDC}\begin{tabular}[c]{@{}c@{}}C (\,nF\,)\end{tabular} & \cellcolor[HTML]{F0DDDC}\begin{tabular}[c]{@{}c@{}}$\hat{E}^r$ (\,$10^{-10}$\,J\,)\end{tabular} & \cellcolor[HTML]{F0DDDC}Freq  \\ \hline
0.25 & 4.32 & 0.22 & 1 & 2.41 & 1.08 \\ \hline
0.5 & 4.21 & 0.61 & 2 & 4.21 & 0.61 \\ \hline
1 & 4.00 & 2.38 & 4 & 9.11 & 0.29 \\ \hline
\end{tabular}
\end{table}

We define the improvement of accuracy as $(E^{es}_{\beta_{i+1}}\!\!-E^{es}_{\beta_i})/E^{es}_{\beta_i}$, where $\beta_0\!=\!5.25\!\times\!10^7$ is the initial grid density and $\beta_i\!=\!i\beta_0$ represents the grid density in the $i^{th}$ iteration. In Fig.~\!\ref{fig:CdfDensity}, we set the threshold of convergence as $0.5\%$ of the accuracy improvement and show the cumulative distribution function (CDF)  of grid density with respect to the packet generation rates, $\lambda$, capacitances, $C$, and the deadline of data transmission, $T$.

From Fig.~\!\ref{fig:CdfDensity}, it can be observed that in DTER, both $\lambda$ and $C$ have a significant impact on the required density of grids to achieve the default accuracy, but the impact of $T$ is negligible. Using CDF $=0.9$ as an example, when $C\!=\!2$\,nF but increase $\lambda$ from $0.25$ to $1$, the required grid density increases from $3.4\!\times\!10^{9}$ to $1.2\!\times\!10^{11}$, as demonstrated in Fig.~\!\ref{fig:CdfDensity_lambda}. When $\lambda\!=\!0.5$ but increase $C$ from $1$\,nF to $4$\,nF, the required grid density decreases from $6.5\!\times\!10^{10}$ to $3.2\!\times\!10^{9}$, as shown in Fig.~\!\ref{fig:CdfDensity_C}. When $C$ and $\lambda$ are fixed but change $T$ from $50$\,s to $100$\,s, the required grid density remains unchanged, which is $9.4\!\times\!10^{9}$, as shown in Fig.~\!\ref{fig:CdfDensity_T}. Therefore, the observations from Fig.~\!\ref{fig:CdfDensity} verify the previous conclusion, i.e., to achieve a certain accuracy, the required density of grid in DTER is proportional to $\lambda$ but inversely proportional to $C$.

 \subsection{Performance of Online Strategy}
\label{subsec:PerOnline}

Now, we evaluate the performance of suboptimal online strategy that is proposed in Section~\!\ref{sec:OnlineOpt} for the second step of DTER. The energy efficiency of the online strategy in a single energy tunnel is assessed first, the results are then extended to a multi-tunnel scenario.

\begin{figure}[htb]
\centerline{\includegraphics[width=7.1cm]{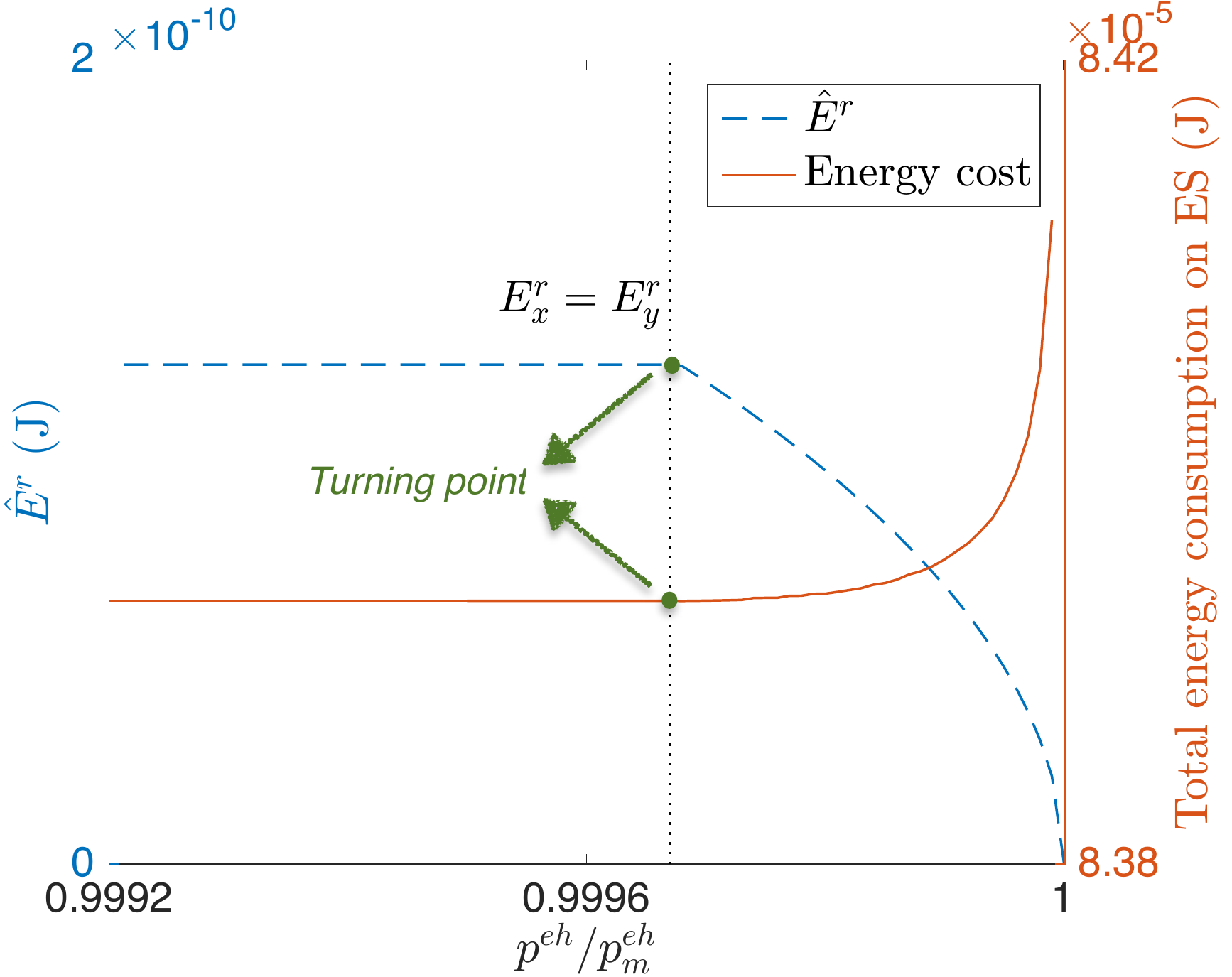}}
  \caption{The optimal request energy, $\hat{E}^r$, and the total energy consumption at ES  with respect to the normalized transmission power, $p^{eh}/p^{eh}_m$,  of an EHD.}\label{fig:charg_cons_p}
\end{figure}

Fig.~\!\ref{fig:charg_cons_p} provides insight into how the EHD's transmission power, $p^{eh}$, affects the optimal energy, $\hat{E}^r$, requested by the EHD each time and the total energy consumed by the ES. Since $\hat{E}^r$ is several orders of magnitude larger than the total energy consumption of the ES, we use different scales to display two curves in the same figure. In the simulation, the EHD is scheduled to harvest a total amount of $8$\,nJ of energy in a single tunnel with different $p^{eh}/p^{eh}_m$. Here, $p^{eh}_m$ is the highest transmission power that an EH system can afford, which is given by (\ref{eq:eh10}), to normalize $p^{eh}$. 

As illustrated in Fig.~\!\ref{fig:charg_cons_p}, if $p^{eh}$ is not extremely close to $p^{eh}_m$, both $\hat{E}^r$ and the total energy consumption of the ES  are constant. However, once $p^{eh}$ approaches $p^{eh}_m$, i.e., $p^{eh}/p^{eh}_m\!\to\!1$, $\hat{E}^r$ reduces from $1.24\!\times\!10^{-10}$\,J to zero, while the overall energy consumption of the ES rises from $8.39\!\times\!10^{-5}$\,J towards infinity quickly. This phenomenon can be interpreted by  the analysis presented in Section~\!\ref{sec:OnlineOpt} that $\hat{E}^r$ is determined by $E^r_x$ and $E^r_y$, where $E^r_x$  and  $E^r_y$ are obtained from the objective function and the charging constraint, respectively. In particular, we mark a turning point of $\hat{E}^r$ and total energy consumption of the ES with respect to the normalize $p^{eh}$ in the figure. This turning point comes from a specific $p^{eh}$, denoted by $p^{eh}_0$ , that makes $E^r_x\!=\!E^r_y$. When $p^{eh}\leq p^{eh}_0$, i.e., before the turning point, the total energy consumption of an ES and $\hat{E}^r$ in a single tunnel remain unchanged regardless of $p^{eh}$. This is because in this situation, $\hat{E}^r\!=\!E^r_x$, and $E^r_x$ is independent of $p^{eh}$ according to (\ref{eq:eh14}) and (\ref{eq:eh15}). If $p^{eh}_0<p^{eh}<p^{eh}_m$, $\hat{E}^r$ is equivalent to $E^r_y$, which decreases monotonically with the increase of normalized $p^{eh}$ based on (\ref{eq:eh16}). In other words, when $p^{eh}\!>\!p^{eh}_0$, i.e., after the turning point, the EHD requests less energy each time but at a higher frequency to harvest sufficient energy in a short period for the EHD running on a high transmission power. However, due to the constant energy paid for each  request, the EHD increases the rate of energy harvesting at the cost of large overhead energy at the ES. When $p^{eh}/p^{eh}_m\!\to\!1$, the charging constraint pushes the EHD to request a tiny energy continuously, and the total energy consumption of the ES goes to infinite.

Next, we study how the length of a feasible energy tunnel, $T$, and the initial energy, $E_0$, in a capacitor affect the energy efficiency of the online strategy. Assume in a single tunnel, the ES consumes a total amount of $E^{es}_{ol}$ and $E^{es}_{op}$ joules of energy with the online and the offline global optimal strategies, respectively. Let $\Delta E$ be the difference between $E^{es}_{ol}$ and $E^{es}_{op}$ in percentage, that is 
\begin{equation}\label{eq:eh21}
    \Delta E=\displaystyle\frac{E^{es}_{ol}-E^{es}_{op}}{E^{es}_{op}}\times100\%.
\end{equation}
In (\ref{eq:eh21}), the energy consumption at ES with the online strategy is equal to that with the global optimal solution if  $\Delta E\!=\!0$. In this case, we claim that the energy efficiency of the online strategy reaches $100\%$. By contrast, if ES consumes much higher energy with the online strategy than  with the global optimal one, $\Delta E$ rises to infinity and the energy efficiency of the online strategy drops to $0$.

\begin{figure}[htb]
\centerline{\includegraphics[width=6.7cm]{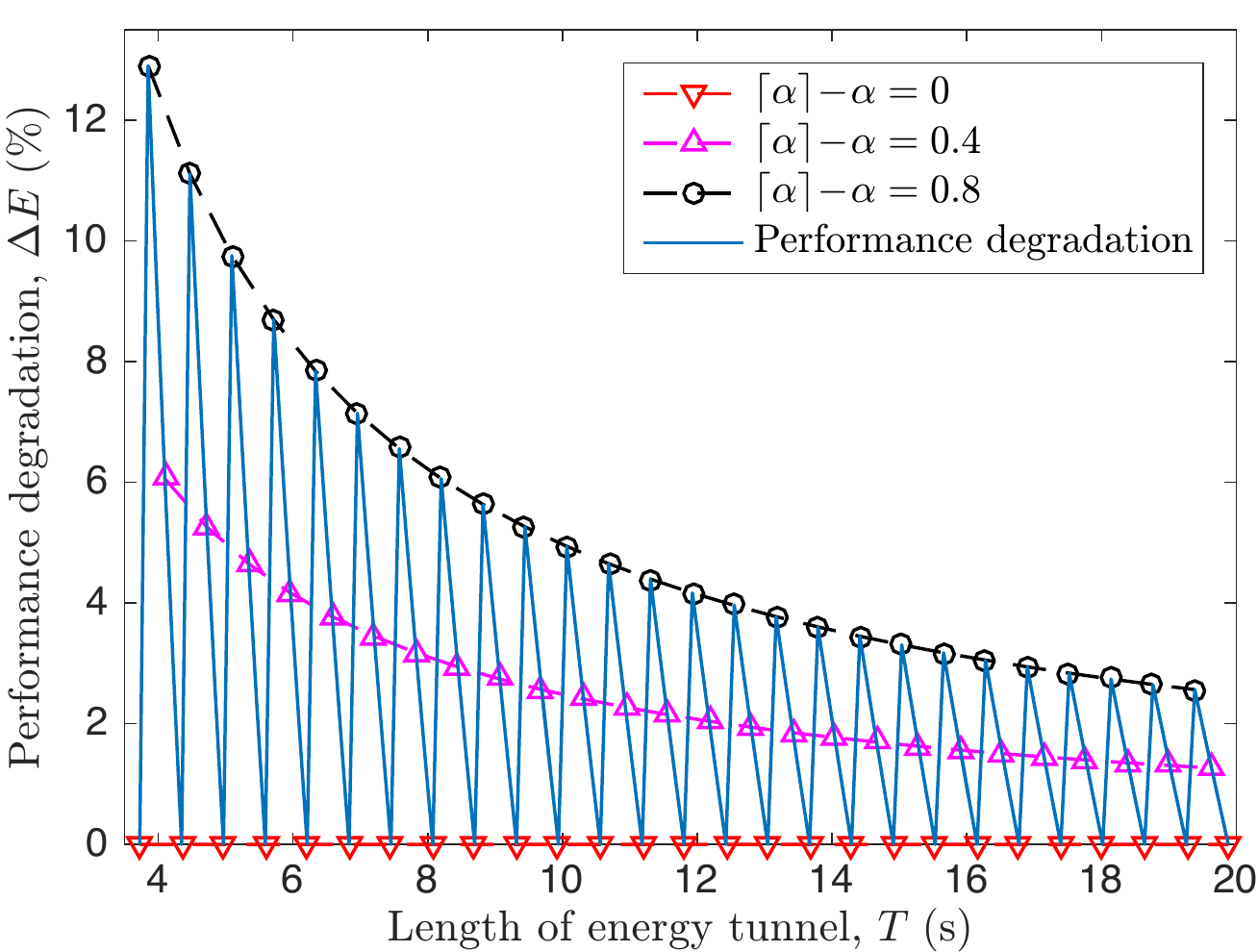}}
  \caption{The degradation of the online strategy's energy efficiency in a single tunnel with respect to $T$.}\label{fig:dppw_cmp_T}
\end{figure}

In Fig.~\!\ref{fig:dppw_cmp_T}, we set $E_0\!=\!\hat{E}^b$ and increase $T$ from $4$\,s to $20$\,s to show how $\Delta E$ changes with the length of energy tunnel, where $\hat{E}^b$ is obtained from (\ref{eq:eh14}). The solid blue curve depicts the degradation of the online strategy's energy efficiency, which periodically reduces versus the length of energy tunnel. It is worthy of pointing out that the performance degradation of online strategy is caused by the inefficiency of its last energy request. To be specific, if the number of requests (i.e., $\alpha$) calculated in (\ref{eq:eh18}) is not an integer, the online strategy will request $\lceil\alpha\rceil$ times, where $\lceil\cdot\rceil$ is the ceiling function. In such an case, the EHD usually harvests an excess of energy from the last request, as the online solution is unaware of the accurate amount of energy needed in the future. When an excessive energy request occurs, the larger the $\lceil\alpha\rceil \!-\!\alpha$, the more wasted energy requested by the online strategy. To demonstrate how $\lceil\alpha\rceil \!-\!\alpha$ affects the performance of the online strategy, we select three cases when $\lceil\alpha\rceil \!-\!\alpha\!=\!0$, $0.4$ and $0.8$. The trend of performance degradation in each case is plotted in Fig.~\!\ref{fig:dppw_cmp_T} with a dot line, from which it can be observed that (a) for $\lceil\alpha\rceil \!-\!\alpha\!=\!0$, the online strategy obtains an appropriate amount of energy from its last request, thereby it achieves the same performance as the global optimal solution; (b) in a same period, i.e., for the same $\alpha$, $\Delta E$ increases with the growth of $\lceil\alpha\rceil \!-\!\alpha$.

Moreover, Fig.~\!\ref{fig:dppw_cmp_T} illustrates that for the same $\lceil\alpha\rceil \!-\!\alpha$, $\Delta E$ reduces as the increase of $T$. This is because (a) $E^{es}_{op}$ rises linearly with $T$, and (b) the online and the offline optimal strategies have almost the same energy efficiency before the last energy request, which implies that the absolute difference of energy efficiency, $E^{es}_{ol}-E^{es}_{op}$, in (\ref{eq:eh21}) changes with $\lceil\alpha\rceil -\alpha$ rather than $T$. Considering the worst case where $\alpha$ slightly exceeds an integer, i.e., $\lceil\alpha\rceil \!-\!\alpha\!\approx\!1$, the online strategy requests $\hat{E}^r$ joules of excessive energy, namely, $E^{es}_{ol}-E^{es}_{op}\approx\!\hat{E}^r$ and $E^{es}_{op}\approx\!(\lceil\alpha\rceil-1)\!\times\!\hat{E}^r$. Eventually, according to (\ref{eq:eh21}), it can be obtained that $\Delta E\!\approx\!\frac{1}{\lceil\alpha\rceil-1}\!\times\!100\%$. Therefore, with the increase of $\alpha$ or $T$, $\Delta E$ decreases to a small value quickly, even in the worst situation. This is verified in Fig.~\!\ref{fig:dppw_cmp_T}, where the amplitude of oscillation of $\Delta E$ on the black curve drops from $12.9\%$ to $2.6\%$ when $T$ increases from $3.9$\,s to $19.4$\,s. In a real application, the operation time of an EH system is usually long, hence the energy efficiency of the online strategy can be very close to the offline global optimal solution. 

\begin{figure}[htb]
\centerline{\includegraphics[width=6.6cm]{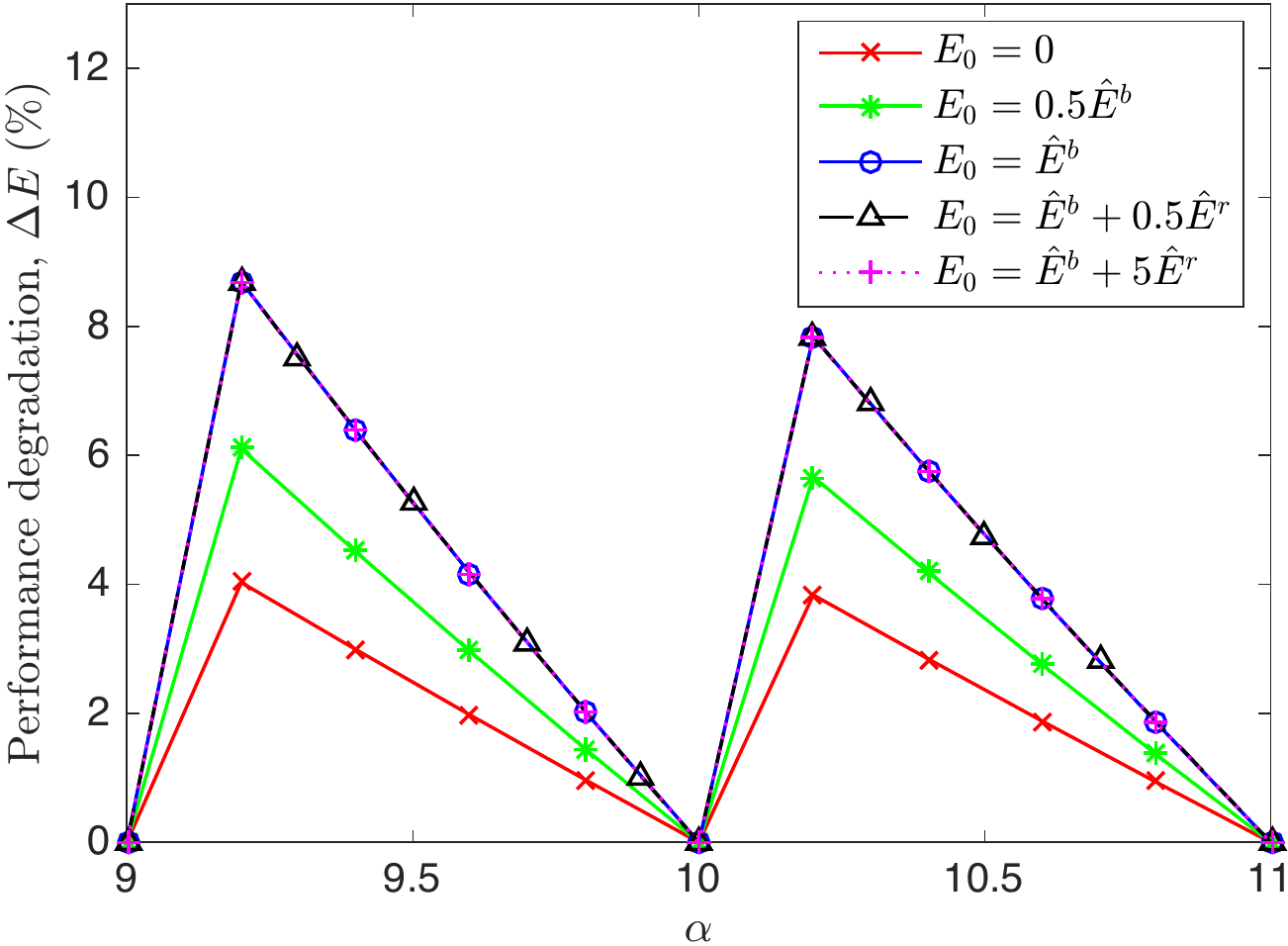}}
  \caption{The degradation of the online strategy's energy efficiency in a single tunnel with different $\alpha$ and $E_0$, where the top three curves (i.e., blue, black and pink) overlapped.}\label{fig:dppw_cmp_alpha}
\end{figure}

In Fig.~\!\ref{fig:dppw_cmp_alpha}, we vary both the number of energy requests, $\alpha$, and the initial energy, $E_0$, to study how the two factors jointly impact the performance of the online strategy.  Next, we summarize our observations into the following three situations.
\vspace{0.1cm}
\begin{adjustwidth}{0cm}{0cm}
    \begin{itemize}
    \vspace{0.12cm}
    \item[(a)] $0\leq E_0\leq\hat{E}^b$: By comparing the red, green and blue curves in Fig.~\!\ref{fig:dppw_cmp_alpha}, we can obtain that the performance degeneration of the online strategy is zero when $\alpha$ is an integer, i.e., $\lceil\alpha\rceil \!-\!\alpha\!=\!0$. This indicates that if the initial energy of an EHD is smaller than $\hat{E}^b$, the online and the optimal strategies perform the same procedure for their first energy request: they both send a request at the beginning of a feasible energy tunnel to charge the capacitor to $\hat{E}^b\!+\!\hat{E}^r$. With such a procedure, $E^{es}_{op}$, i.e., the denominator of (\ref{eq:eh21}), increases with the decrease of $E_0$; however, since $E^{es}_{ol}-E^{es}_{op}$, i.e., the numerator of (\ref{eq:eh21}),  is independent of $E_0$. Therefore, $\Delta E$ is inversely proportional to $E_0$. In Fig.~\!\ref{fig:dppw_cmp_alpha}, it is proved by comparing the blue, red and green curves at the same $\alpha$, where $\Delta E$ grows with the increase of $E_0$.
    \vspace{0.12cm}
    \item[(b)] $\hat{E}^b\!<\!E_0\!<\!\hat{E}^b\!+\!\hat{E}^r$: In this case, the online method sends no energy request until the residual energy in a capacitor drops to $\hat{E}^b$. For the global optimal strategy, the EHD has two potential actions: (a) sending the initial request when residual energy drops to $\hat{E}^b$, which is the same as the online strategy, or (b) sending the initial request at the beginning of the  tunnel to rise the energy of the capacitor to $\hat{E}^b\!+\!\hat{E}^r$. We testify that the offline and online strategies take the same action for the initial request through simulations, which implies that during the first $(E_0-\hat{E}^b)/p^{eh}$ seconds, $\Delta E$ in (\ref{eq:eh21}) is zero. At the instant of initial energy request, the residual energy of both online and global optimal strategies starts from $\hat{E}^b$, and the trend of $\Delta E$ with respect to $\alpha$ is the same as Fig.~\!\ref{fig:dppw_cmp_T}. Therefore, when $\hat{E}^b\!\leq\!E_0\!\leq\!\hat{E}^b\!+\!\hat{E}^r$, $\Delta E$ is independent of $E_0$ as revealed in Fig.~\!\ref{fig:dppw_cmp_alpha}, where the black curve overlaps the blue one.
     \vspace{0.12cm}
    \item[(c)] $\hat{E}^b\!+\!\hat{E}^r \leq  E_0 \leq E^{eh}_m$: Both the offline global optimal and online strategies in this situation use the initial energy first, and then initiate the first request when the residual energy reduces to $\hat{E}^b$. Therefore, the profile of $\Delta E$ with respect to $\alpha$ is the same as the case (b), which causes the pink curve in Fig.~\!\ref{fig:dppw_cmp_alpha} to overlapping the black one.
 \end{itemize}
 \end{adjustwidth}
\vspace{0.1cm}

To summarize, $\Delta E$ changes periodically with $\alpha$, and the highest degradation in a period decreases with the increase of $\lceil\alpha\rceil$. Moreover, when $E_0\leq\hat{E}^b$, $\Delta E$ is inversely proportional to $E_0$; when $\hat{E}^b< E_0\leq E^{eh}_m$, $\Delta E$ is independent of $E_0$.

\begin{figure}[htb]
\centerline{\includegraphics[width=6.8cm]{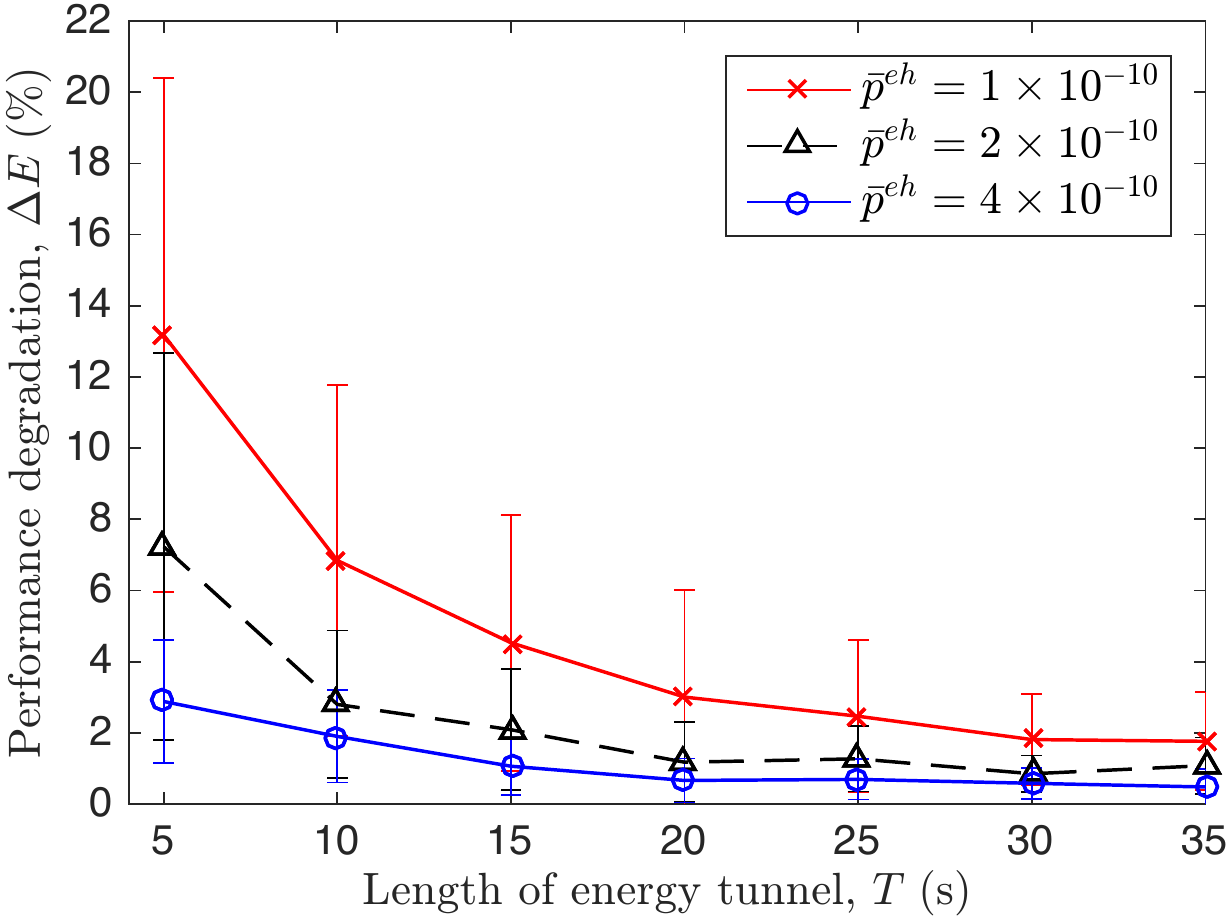}}
  \caption{Comparison of energy efficiencies between the global optimal and online strategies in multiple tunnels with different $\bar{T}$ and $\bar{p}^{eh}$.}\label{fig:dppw_cmp_errbar}
\end{figure}

In Fig.~\!\ref{fig:dppw_cmp_errbar}, we compare the online and the offline optimal solution in multi-tunnel tests. In each test, the  feasible energy tunnel consists of three pieces. The length and the slope (e.g., transmission power of an EHD) of each piece are random, which follow a Gaussian distribution. Moreover, the standard variation of the slope and that of the length is set equivalent to their mean value. The average transmission power, $\bar{p}^{eh}$, of an EHD ranges from $0.1$\,nW to $0.4$\,nW and the overall length, $T$, of the multi-tunnel varies between $5$\,s and $35$\,s.

Similar to the single tunnel scenario, $\Delta E$ in the multiple tunnels oscillates with $\lceil\alpha\rceil \!-\!\alpha$, but the highest amplitude reduces with the length of energy tunnel in the long term. This is verified in Fig.~\!\ref{fig:dppw_cmp_errbar}, which demonstrates the monotonic decrease of the average $\Delta E$ with $T$. Recall that the main reason causing the difference of energy efficiency between the online and global optimal strategies is the potential superfluous energy harvested occurred in the last energy request. With the increase of tunnel length, the ratio of wasted energy to the total harvested energy reduces, which results in less performance degradation. For the similar reason, we can also observe the  difference of energy efficiency decreases with the growth of $\bar{p}^{eh}$. With a larger $\bar{p}^{eh}$, the amount of energy to replenish is more with a fixed length of energy tunnel and percentage of the wasted energy in turn is reduced. Therefore, we can expect comparable performance of the online strategy with the global optimal solution in the long term.

\section{Conclusions}
\label{sec:conclusion}

In this paper, we have presented an optimal energy requesting strategy, termed dual tunnel
energy requesting (DTER), for RF energy harvesting IoT.
The key feature of DTER is that both the nonlinear charging feature
and the overhead issue are taken into account to minimize the overall
energy consumption. We have investigated both offline and online scenarios for DTER and evaluated them through theoretical analysis and simulation study. Simulation results have demonstrated that the  online strategy achieves comparable energy efficiency with the offline optimal solution in a long term. This paper provides a lower bound of the overall energy consumption at both energy harvesting device and energy source. The research of this paper is expected to shed light on the future research of energy harvesting communications.

\appendix
\section{Appendix}
\label{app:appendix}

\subsection{Proof of Lemma~\!\ref{lem:06}}
\label{app:appA}

\begin{proof}
	After $t_p$, denote the overall energy consumption of an EHD with a transmission rate $r$ as $E(r)$. Let $r^{eh}_1$ and $r^{eh}_2$ be two transmission rates that $r^{eh}_1$ is larger than zero and $r^{eh}_2$ approaches zero, i.e., $r^{eh}_1\!>\!0$ and $r^{eh}_2\!\to\!0$. According to (\ref{eq:eh04}), we have that 
	\begin{equation}\label{eq:eh22}
	\begin{array}{lll}
	\vspace{0.1cm}
	E(r^{eh}_1)\!\!\!\!\!\!&-&\!\!\!\!\!E(r^{eh}_2)\\
	\vspace{0.1cm}
		\!\!\!\!\!\!&=&\!\!\!\!\displaystyle\frac{f(r^{eh}_1)D_m^{eh}}{r^{eh}_1}-\lim_{r^{eh}_2 \to 0}\frac{f(r^{eh}_2)D_m^{eh}}{r^{eh}_2}\\
	\vspace{0.1cm}
											\!\!\!\!\!&=&\!\!\!\!\!\displaystyle\lim_{r^{eh}_2 \to 0}\!\!\frac{\left[f(r^{eh}_1)-r^{eh}_1f^\prime(r^{eh}_2)\right]D_m^{eh}}{r^{eh}_1}\\
    \vspace{0.1cm}
											\!\!\!\!\!&=&\!\!\!\!\!\displaystyle\lim_{r^{eh}_2 \to 0}\!\!\frac{\left[f(r^{eh}_1)\!-\!f(r^{eh}_2)\!-\!(r^{eh}_1\!-\!r^{eh}_2)f^\prime(r^{eh}_2)\right]D_m^{eh}}{r^{eh}_1}\\
											\!\!\!\!\!&>&\!\!\!\!\!0,
	\end{array}
	\end{equation}
	where $f^\prime(\cdot)$ represents the differentiation of $f(\cdot)$. The expression of $E(r^{eh}_2)$ involves $0/0$. Since $f(\cdot)$ is continuously differentiable, the L'H\^opital's rule is applied to get the second equation of (\ref{eq:eh22}). Additionally, the derivation of the last inequality is based on the property of a strictly convex function that $f(x)-f(y)\!>\!f^\prime(y)(x-y)$.
	From (\ref{eq:eh22}), it can be obtained that $\forall\, r^{eh}_1\!>\!0, r^{eh}_2\!\to\!0\!: E(r^{eh}_2)\!<\!E(r^{eh}_1)$. Accordingly, the optimal transmission rate approaches zero after $t_p$.
\end{proof}

\subsection{Notations}
\label{app:appB}
We list the definitions of all symbols used below.
\begin{table}[h]
\centering
\footnotesize
\label{tab:Notation}
\begin{tabular}{ll}

\hline

\multicolumn{1}{|c|}{\cellcolor[HTML]{FFCB2F}Symbol} & \multicolumn{1}{c|}{\cellcolor[HTML]{FFCB2F}Definition} \\ \hline\hline

\multicolumn{1}{|c|}{\cellcolor[HTML]{BBDAFF}$\alpha$}  & \multicolumn{1}{l|}{\cellcolor[HTML]{BBDAFF}Number of requests in a piece of energy tunnel} \\ \hline

\multicolumn{1}{|c|}{\cellcolor[HTML]{ECF4FF}$\hat{\alpha}$}  & \multicolumn{1}{l|}{\cellcolor[HTML]{ECF4FF}Optimal $\alpha$ in a piece of energy tunnel} \\ \hline

\multicolumn{1}{|c|}{\cellcolor[HTML]{BBDAFF}$C$}  & \multicolumn{1}{l|}{\cellcolor[HTML]{BBDAFF}Capacitance of the charging circuit} \\ \hline

\multicolumn{1}{|c|}{\cellcolor[HTML]{ECF4FF}$D^{eh}_m$}  & \multicolumn{1}{l|}{\cellcolor[HTML]{ECF4FF}Capacity of data storage on EHD} \\ \hline

\multicolumn{1}{|c|}{\cellcolor[HTML]{BBDAFF}$D^{b}_{t}$}  & \multicolumn{1}{l|}{\cellcolor[HTML]{BBDAFF}Residual data in storage on EHD at time $t$} \\ \hline

\multicolumn{1}{|c|}{\cellcolor[HTML]{ECF4FF}$D_n$}& \multicolumn{1}{l|}{\cellcolor[HTML]{ECF4FF}Size of  data packet, $n$} \\ \hline

\multicolumn{1}{|c|}{\cellcolor[HTML]{BBDAFF}$e_r$}& \multicolumn{1}{l|}{\cellcolor[HTML]{BBDAFF}Overhead for energy request} \\ \hline

\multicolumn{1}{|c|}{\cellcolor[HTML]{ECF4FF}$E^{es}_i$}& \multicolumn{1}{l|}{\cellcolor[HTML]{ECF4FF}Energy cost on ES of the $i^{th}$ energy transfer} \\ \hline

\multicolumn{1}{|c|}{\cellcolor[HTML]{BBDAFF}$E^{es}_{ol/op}$}& \multicolumn{1}{l|}{\cellcolor[HTML]{BBDAFF}Energy consumption by ES with online/optimal strategy} \\ \hline


\multicolumn{1}{|c|}{\cellcolor[HTML]{ECF4FF}$E^r_i$}  & \multicolumn{1}{l|}{\cellcolor[HTML]{ECF4FF}Harvested energy from energy packet, $i$} \\ \hline

\multicolumn{1}{|c|}{\cellcolor[HTML]{BBDAFF}$E^{eh}_m$}  & \multicolumn{1}{l|}{\cellcolor[HTML]{BBDAFF}Energy capacity of the battery on EHD} \\ \hline

\multicolumn{1}{|c|}{\cellcolor[HTML]{ECF4FF}$E^b_{t_{r_i}}$}& \multicolumn{1}{l|}{\cellcolor[HTML]{ECF4FF}Residual energy at time $t_{r_i}$} \\ \hline

\multicolumn{1}{|c|}{\cellcolor[HTML]{BBDAFF}$E_0$}& \multicolumn{1}{l|}{\cellcolor[HTML]{BBDAFF}Initial energy on EHD} \\ \hline

\multicolumn{1}{|c|}{\cellcolor[HTML]{ECF4FF}$f(\cdot)$}& \multicolumn{1}{l|}{\cellcolor[HTML]{ECF4FF}Rate-power function, an inverse function of $g(\cdot)$} \\ \hline

\multicolumn{1}{|c|}{\cellcolor[HTML]{BBDAFF}$g(\cdot)$}& \multicolumn{1}{l|}{\cellcolor[HTML]{BBDAFF}Power-rate function, an inverse function of $f(\cdot)$} \\ \hline

\multicolumn{1}{|c|}{\cellcolor[HTML]{ECF4FF}$p^{c}$}& \multicolumn{1}{l|}{\cellcolor[HTML]{ECF4FF}Charging rate on the EHD} \\ \hline

\multicolumn{1}{|c|}{\cellcolor[HTML]{BBDAFF}$p^{eh}$}& \multicolumn{1}{l|}{\cellcolor[HTML]{BBDAFF}Transmission power of EHD} \\ \hline

\multicolumn{1}{|c|}{\cellcolor[HTML]{ECF4FF}$p^{es}$}& \multicolumn{1}{l|}{\cellcolor[HTML]{ECF4FF}Transmission power of ES} \\ \hline

\multicolumn{1}{|c|}{\cellcolor[HTML]{BBDAFF}$p^{eh}_m/o$}& \multicolumn{1}{l|}{\cellcolor[HTML]{BBDAFF}Maximum/optimal harvested power on the EHD} \\ \hline


\multicolumn{1}{|c|}{\cellcolor[HTML]{ECF4FF}$r^{eh}$} & \multicolumn{1}{l|}{\cellcolor[HTML]{ECF4FF}Transmission rate of EHD} \\ \hline

\multicolumn{1}{|c|}{\cellcolor[HTML]{BBDAFF}$r^{eh}_o$} & \multicolumn{1}{l|}{\cellcolor[HTML]{BBDAFF}Optimal transmission rate of EHD} \\ \hline

\multicolumn{1}{|c|}{\cellcolor[HTML]{ECF4FF}$R$}  & \multicolumn{1}{l|}{\cellcolor[HTML]{ECF4FF}Resistance of the charging circuit} \\ \hline

\multicolumn{1}{|c|}{\cellcolor[HTML]{BBDAFF}$t_{d_n}$} & \multicolumn{1}{l|}{\cellcolor[HTML]{BBDAFF}Arrival time of data packet, $d$} \\ \hline 
       
\multicolumn{1}{|c|}{\cellcolor[HTML]{ECF4FF}$t_{r_i}$} & \multicolumn{1}{l|}{\cellcolor[HTML]{ECF4FF}Arrival time of energy packet, $i$} \\ \hline    
         
\multicolumn{1}{|c|}{\cellcolor[HTML]{BBDAFF}$T$} & \multicolumn{1}{l|}{\cellcolor[HTML]{BBDAFF}End time of the data tunnel} \\ \hline        

\multicolumn{1}{|c|}{\cellcolor[HTML]{ECF4FF}$T^{es}_i$} & \multicolumn{1}{l|}{\cellcolor[HTML]{ECF4FF}Length of energy packet, $i$} \\ \hline


\multicolumn{1}{|c|}{\cellcolor[HTML]{BBDAFF}$V_{j,k}$} & \multicolumn{1}{l|}{\cellcolor[HTML]{BBDAFF}Vertex with coordinates ($j,k$)} \\ \hline

\multicolumn{1}{|c|}{\cellcolor[HTML]{ECF4FF}$V_m$} & \multicolumn{1}{l|}{\cellcolor[HTML]{ECF4FF}Maximum voltage that EHD can reach} \\ \hline


\multicolumn{1}{|c|}{\cellcolor[HTML]{BBDAFF}$w_{(j,k)\!\to\!(m,n)}$} & \multicolumn{1}{l|}{\cellcolor[HTML]{BBDAFF}Weight (i.e., energy cost) of edge $(j, k)\to(m, n)$} \\ \hline

\multicolumn{1}{|c|}{\cellcolor[HTML]{ECF4FF}$z(\cdot)$} & \multicolumn{1}{l|}{\cellcolor[HTML]{ECF4FF}Charging function} \\ \hline
\end{tabular}
\end{table}



\begin{thebibliography}{10}
\providecommand{\url}[1]{#1}
\csname url@samestyle\endcsname
\providecommand{\newblock}{\relax}
\providecommand{\bibinfo}[2]{#2}
\providecommand{\BIBentrySTDinterwordspacing}{\spaceskip=0pt\relax}
\providecommand{\BIBentryALTinterwordstretchfactor}{4}
\providecommand{\BIBentryALTinterwordspacing}{\spaceskip=\fontdimen2\font plus
\BIBentryALTinterwordstretchfactor\fontdimen3\font minus
  \fontdimen4\font\relax}
\providecommand{\BIBforeignlanguage}[2]{{%
\expandafter\ifx\csname l@#1\endcsname\relax
\typeout{** WARNING: IEEEtran.bst: No hyphenation pattern has been}%
\typeout{** loaded for the language `#1'. Using the pattern for}%
\typeout{** the default language instead.}%
\else
\language=\csname l@#1\endcsname
\fi
#2}}
\providecommand{\BIBdecl}{\relax}
\BIBdecl

\bibitem{gubbi2013internet}
J.~Gubbi, R.~Buyya, S.~Marusic, and M.~Palaniswami, ``Internet of things (iot):
  A vision, architectural elements, and future directions,'' \emph{Future
  generation computer systems}, vol.~29, no.~7, pp. 1645--1660, 2013.

\bibitem{kamalinejad2015wireless}
P.~Kamalinejad, C.~Mahapatra, Z.~Sheng, S.~Mirabbasi, V.~C. Leung, and Y.~L.
  Guan, ``Wireless energy harvesting for the internet of things,'' \emph{IEEE
  Communications Magazine}, vol.~53, no.~6, pp. 102--108, 2015.

\bibitem{kazmierski2014energy}
T.~J. Kazmierski and S.~Beeby, \emph{Energy harvesting systems}.\hskip 1em plus
  0.5em minus 0.4em\relax Springer, 2014.

\bibitem{ulukus2015energy}
S.~Ulukus, A.~Yener, E.~Erkip, O.~Simeone, M.~Zorzi, P.~Grover, and K.~Huang,
  ``{Energy harvesting wireless communications: a review of recent advances},''
  \emph{IEEE Journal on Selected Areas in Communications}, vol.~33, no.~3, pp.
  360--381, 2015.

\bibitem{zhou2015greendelivery}
S.~Zhou, J.~Gong, Z.~Zhou, W.~Chen, and Z.~Niu, ``{GreenDelivery: proactive
  content caching and push with energy-harvesting-based small cells},''
  \emph{IEEE Communications Magazine}, vol.~53, no.~4, pp. 142--149, 2015.

\bibitem{pinuela2013ambient}
M.~Pi{\~n}uela, P.~D. Mitcheson, and S.~Lucyszyn, ``{Ambient RF energy
  harvesting in urban and semi-urban environments},'' \emph{IEEE Transactions
  on Microwave Theory and Techniques}, vol.~61, no.~7, pp. 2715--2726, 2013.

\bibitem{liu2013ambient}
V.~Liu, A.~Parks, V.~Talla, S.~Gollakota, D.~Wetherall, and J.~R. Smith,
  ``{Ambient backscatter: wireless communication out of thin air},'' in
  \emph{Proceedings of SIGCOMM}.\hskip 1em plus 0.5em minus 0.4em\relax ACM,
  2013, pp. 39--50.

\bibitem{luo2017optimal}
Y.~Luo, L.~Pu, Y.~Zhao, G.~Wang, and M.~Song, ``{Optimal energy requesting
  strategy for RF-based energy harvesting wireless communications},'' in
  \emph{Proceedings of INFOCOM}.\hskip 1em plus 0.5em minus 0.4em\relax IEEE,
  2017, pp. 1--9.

\bibitem{mishra2015smart}
D.~Mishra, S.~De, S.~Jana, S.~Basagni, K.~Chowdhury, and W.~Heinzelman,
  ``{Smart RF energy harvesting communications: challenges and
  opportunities},'' \emph{IEEE Communications Magazine}, vol.~53, no.~4, pp.
  70--78, 2015.

\bibitem{kaushik2013experimental}
K.~Kaushik, D.~Mishra, S.~De, S.~Basagni, W.~Heinzelman, K.~Chowdhury, and
  S.~Jana, ``{Experimental demonstration of multi-hop RF energy transfer},'' in
  \emph{Proceedings of the International Symposium on Personal Indoor and
  Mobile Radio Communications}.\hskip 1em plus 0.5em minus 0.4em\relax IEEE,
  2013, pp. 538--542.

\bibitem{huang2014enabling}
K.~Huang and V.~K. Lau, ``{Enabling wireless power transfer in cellular
  networks: architecture, modeling and deployment},'' \emph{IEEE Transactions
  on Wireless Communications}, vol.~13, no.~2, pp. 902--912, 2014.

\bibitem{yang2012optimal}
J.~Yang and S.~Ulukus, ``{Optimal packet scheduling in an energy harvesting
  communication system},'' \emph{IEEE Transactions on Communications}, vol.~60,
  no.~1, pp. 220--230, 2012.

\bibitem{tutuncuoglu2012optimum}
K.~Tutuncuoglu and A.~Yener, ``{Optimum transmission policies for battery
  limited energy harvesting nodes},'' \emph{IEEE Transactions on Wireless
  Communications}, vol.~11, no.~3, pp. 1180--1189, 2012.

\bibitem{lu2015resource}
X.~Lu, P.~Wang, D.~Niyato, and Z.~Han, ``{Resource allocation in wireless
  networks with RF energy harvesting and transfer},'' \emph{IEEE Network},
  vol.~29, no.~6, pp. 68--75, 2015.

\bibitem{das2014rfid}
R.~Das and P.~Harrop, ``{RFID forecasts, players and opportunities
  2014--2024},'' \emph{IDTechEx report}, 2014.

\bibitem{kaur2011rfid}
M.~Kaur, M.~Sandhu, N.~Mohan, and P.~S. Sandhu, ``{RFID technology principles,
  advantages, limitations \& its applications},'' \emph{International Journal
  of Computer and Electrical Engineering}, vol.~3, no.~1, pp. 151--157, 2011.

\bibitem{percy2012supplying}
S.~Percy, C.~Knight, F.~Cooray, and K.~Smart, ``{Supplying the power
  requirements to a sensor network using radio frequency power transfer},''
  \emph{Sensors}, vol.~12, no.~7, pp. 8571--8585, 2012.

\bibitem{huang2015cutting}
K.~Huang and X.~Zhou, ``{Cutting the last wires for mobile communications by
  microwave power transfer},'' \emph{IEEE Communications Magazine}, vol.~53,
  no.~6, pp. 86--93, 2015.

\bibitem{lu2015wireless}
X.~Lu, P.~Wang, D.~Niyato, D.~I. Kim, and Z.~Han, ``{Wireless networks with RF
  energy harvesting: a contemporary survey},'' \emph{IEEE Communications
  Surveys \& Tutorials}, vol.~17, no.~2, pp. 757--789, 2015.

\bibitem{naderi2014rf}
M.~Y. Naderi, P.~Nintanavongsa, and K.~R. Chowdhury, ``{RF-MAC: A medium access
  control protocol for re-chargeable sensor networks powered by wireless energy
  harvesting},'' \emph{IEEE Transactions on Wireless Communications}, vol.~13,
  no.~7, pp. 3926--3937, 2014.

\bibitem{ozel2011transmission}
O.~Ozel, K.~Tutuncuoglu, J.~Yang, S.~Ulukus, and A.~Yener, ``{Transmission with
  energy harvesting nodes in fading wireless channels: optimal policies},''
  \emph{IEEE Journal on Selected Areas in Communications}, vol.~29, no.~8, pp.
  1732--1743, 2011.

\bibitem{gorlatova2013networking}
M.~Gorlatova, A.~Wallwater, and G.~Zussman, ``{Networking low-power energy
  harvesting devices: Measurements and algorithms},'' \emph{IEEE Transactions
  on Mobile Computing}, vol.~12, no.~9, pp. 1853--1865, 2013.

\bibitem{nintanavongsa2012design}
P.~Nintanavongsa, U.~Muncuk, D.~R. Lewis, and K.~R. Chowdhury, ``{Design
  optimization and implementation for RF energy harvesting circuits},''
  \emph{IEEE Journal on Emerging and Selected Topics in Circuits and Systems},
  vol.~2, no.~1, pp. 24--33, 2012.

\bibitem{sniedovich2010dynamic}
M.~Sniedovich, \emph{{Dynamic programming: foundations and principles}}.\hskip
  1em plus 0.5em minus 0.4em\relax CRC press, 2010.

\end{thebibliography}

\end{document}